\documentclass[a4paper,11pt]{article}
\pdfoutput=1 % if your are submitting a pdflatex (i.e. if you have
\usepackage{jcappub} % for details on the use of the package, please
\usepackage[T1]{fontenc} % if needed
\usepackage{footnote}
\usepackage[latin9]{inputenc}
\usepackage[active]{srcltx}
\usepackage{color}
\usepackage{wasysym}
\usepackage{graphicx}
\usepackage{esint}

\makeatletter

\providecommand{\tabularnewline}{\\}

 \@ifundefined{textcolor}{}
 {%
   \definecolor{BLACK}{gray}{0}
   \definecolor{WHITE}{gray}{1}
   \definecolor{RED}{rgb}{1,0,0}
   \definecolor{GREEN}{rgb}{0,1,0}
   \definecolor{BLUE}{rgb}{0,0,1}
   \definecolor{CYAN}{cmyk}{1,0,0,0}
   \definecolor{MAGENTA}{cmyk}{0,1,0,0}
   \definecolor{YELLOW}{cmyk}{0,0,1,0}
 }

\usepackage{aas_macros}

\usepackage{array}
\usepackage{multirow}

\makeatother

\begin{document}

\title{Model-independent constraints on modified gravity from current data and from the Euclid and SKA future surveys.}

\author{Laura Taddei, Matteo Martinelli, Luca Amendola}

\affiliation{Institut f\"ur Theoretische Physik, Ruprecht-Karls-Universit\"at
Heidelberg, Philosophenweg 16, 69120 Heidelberg, Germany.}

\emailAdd{taddei@thphys.uni-heidelberg.de}
\emailAdd{martinelli@lorentz.leidenuniv.nl}
\emailAdd{amendola@thphys.uni-heidelberg.de}

\abstract{
The aim of this paper is to constrain modified gravity with redshift
space distortion observations and supernovae measurements. Compared with a standard 
$\Lambda$CDM analysis, we include three additional free parameters,
namely the initial conditions of the matter perturbations, the overall perturbation normalization, and a scale-dependent
modified gravity parameter modifying the Poisson equation, in an
attempt to perform a more model-independent analysis. First, we constrain
the Poisson parameter $Y$ (also called $G_{\rm eff}$) by using currently
available $f\sigma_{8}$ data and the recent SN catalog JLA.
We find that the inclusion of the  additional free parameters makes
the constraints significantly weaker than when fixing them to the
standard cosmological value. Second, we forecast future constraints on $Y$ by using the predicted
growth-rate data for Euclid and SKA missions. Here again
we point out the weakening of the constraints when the additional
parameters are included. Finally, we adopt as modified gravity Poisson
parameter the specific Horndeski form, and use scale-dependent forecasts
to build an exclusion plot for the Yukawa potential akin to the ones
realized in laboratory experiments, both for the Euclid and the
SKA surveys. 
}
\maketitle

\section{Introduction}

Since its discovery in 1998 \cite{Riess:1998cb,Perlmutter:1998np},
the accelerated expansion of the Universe has remained an open problem
of cosmology, as our General Relativity based description of the Universe
only allows for deceleration unless some additional components, such
as a cosmological constant $\Lambda$ or additional fundamental fields,
are added to the theory. The additional fields might also modify the
effective gravitational potential and give rise to so-called ``modified
gravity'' (MG) models. This has prompted cosmologists to investigate
many alternatives to General Relativity (see e.g. \cite{Amendola2010}
for an overview of currently available models). Current and upcoming
cosmological surveys are finally reaching a sensitivity which will
allow to test modifications of gravity at cosmological scales and
possibly to distinguish these from the simple scenario of a constant
energy density ($\Lambda$CDM model) or uncoupled and unclustered
dark energy.

In general, at linear perturbation level, a modification of scalar
perturbations with respect to the $\Lambda$CDM model can be described
by two functions in the Einstein equations, which generally depend
on time and space or, in Fourier space, on the wavenumber $k$ (see
for instance \cite{Euclid_TWG}). One function is $Y(t,k)$ which
modifies the standard Poisson equation (and also Newton's constant,
so it is often denoted as $G_{eff}$: this notation however can be
confusing since if $Y$ is scale dependent, the expression for $G_{eff}$
in the potential and in the force equations are different). The other
one is the anisotropic stress $\eta(t,k)$, which is the ratio between
the two linear gravitational potentials $\Phi$ and $\Psi$ which
enter the spatial and temporal part, respectively, of the perturbed
Friedmann-Robertson-Walker (FRW) metric. (Note that our perturbation
variables are to be thought of as the root mean square of the corresponding
random variable, so they are positive defined.) In standard gravity,
we have $Y=\eta=1$.

For a non-relativistic perfect fluid, the anisotropic stress is not sourced
by matter at the linear level, so in this case $\eta$ is a genuine indicator of
modified gravity \cite{2013PhRvD..87b3501A}. In \cite{Amendola:2013qna},
the authors employed a model-independent approach to measure $\eta$ without
any assumption on the initial spectrum of perturbations and the bias,
finding a forecasted   error of a few percent if $\eta$ is assumed
constant. On the other hand, $Y\not=$1 can in principle
arise also due to clustering of dark energy, rather than to a modification
of gravity, but if the dark energy field is not coupled to matter
the clustering on sub-horizon scales is expected to be negligible
unless the field mass is so large that the field behaves as dark matter
and therefore does not drive acceleration, or the sound speed is extremely
small (orders of magnitudes smaller than the velocity of light). In
any case, the detection of a deviation of $Y$ from unity would signal
an important manifestation of new physics.

In a large class of scalar field models, the so-called Horndeski Lagrangian,
the two functions $Y,\eta$ take a particular simple form in the quasi-static
limit, i.e. when we neglect the time-derivatives of the perturbation
variables (see next section). This is a good approximation for scales
much below the sound horizon, see e.g. \cite{DeFelice:2011hq,Amendola:2012ky,Silvestri:2013ne}.
For our purposes we assume this approximation to hold;
however, it is worth stressing that this point should be
further addressed, specially given the large scales probed by the survey 
we will discuss in the rest of the paper (see e.g. \cite{Sawicki:2015zya}).
In real space, the Horndeski form gives rise to the Yukawa correction
to the Newtonian potential. This paper is devoted to constraining
the quasi-static scale-dependent form for the Yukawa parameters through
current and future redshift space distortion (RSD) and supernovae data.

The parameter $Y$ enters the equation for the growth of matter perturbation
density contrast $\delta_{m}$. By comparing the theoretical prediction
of the growth of $\delta_{m}$ with the observations provided by the
redshift distortion quantity $f\sigma_{8}(z)$ one can constrain the
modified gravity parameters. However, one needs also to assume an
initial condition for $\delta$ in order to integrate the differential
equation. Almost universally, the choice has been to assume a standard
matter-dominated cosmology at high redshift. This choice, however,
is a very special one and is not justified neither by observations
nor by theory. On the observational side, there is no direct information
on the growth of matter at any epoch between the cosmic microwave
background (CMB) at $z\approx1000$ and $z\approx1$, but only integral
information like the integrated Sachs-Wolfe effect or the CMB lensing.
On the theory side, any model in which dark energy is not negligible
during the matter era will introduce a deviation from standard initial
conditions that should be taken into account (see discussion in \cite{Taddei:2014wqa}).

A further problem arises when using the $f\sigma_{8}(z)$ data. In
fact, the quantity that is observed is the redshift-distorted galaxy
power spectrum, namely 
\begin{equation}
P_{g}(z,k,\mu)=(1+\frac{f(z,k)}{b(z,k)}\mu^{2})^{2}G^{2}(z,k)\sigma_{8}^{2}b^{2}(z,k)P_{m}(k)
\end{equation}
where $f=d\log\delta_{m}/d\log a$ is the growth rate, $b(z,k)$ is
the bias function, $G(z,k)=\delta_{m}(z.k)/\delta_{m}(0,k)$ is the
growth function, $\mu$ is the direction cosine between the line of
sight and the wave vector $\vec{k}$, and $P_{m}=\delta_{m}(0,k)^{2}$
is the present matter power spectrum. Taking $P_{g}^{1/2}(z,k,1)-P_{g}^{1/2}(z,k,0)$
one gets 
\begin{equation}
f(z,k)G(z,k)\sigma_{8}P_{m}^{1/2}(k)=f\sigma_{8}(z)\times P_{m}^{1/2}(k)=\sigma_{8}\delta_{m}'
\end{equation}
(the prime denotes derivative with respect to $\log a$ and $\sigma_8(z)\equiv \sigma_8 G)$. This is
the raw observed quantity. In order to obtain $f\sigma_{8}(z)$ one
needs a model for $P_{m}^{1/2}(k)$ and at this point the assumption
of a standard power spectrum shape is inserted in the procedure. The
initial power spectrum shape can be measured or constrained through
CMB observations but nothing guarantees that the shape has not changed
after recombination, unless we assume again a standard matter era.
The two problems just mentioned can be both seen as initial conditions
problems: the equation for $\delta_{m}$ is second order (see next
section) so we need two free initial conditions, $\delta_{in}$ and
$\delta_{in}'$, in order to carry out a more model-independent estimation
of modified gravity parameters. Of course one is perfectly entitled
to stick with a standard matter era throughout, but one should be
aware that in this case one is testing a combination of standard assumptions
and new physics, and the constraints on the MG parameters will change
when assuming a different matter era. The purpose of this paper is
to investigate possible constraints from future surveys, when these 
common assumptions are removed, thus performing a rather model independent
forecast on general MG theories.

While at first we explore the simplest case in which $Y$ is constant
both in space and time, we then move to the Horndeski scale-dependent
form, still freezing the time behavior. In this way we can constrain
the parameters that enter the Yukawa potential and build a forecast
exclusion plot similar to the one realized in laboratory or within
the solar system.
We follow here the same approach employed in \cite{Taddei:2014wqa}, 
extending the analysis performed there to more general expansion histories,
in order to take into account models which deviate from $\Lambda$CDM also
at the background level. For this purpose we use a CPL form \cite{Chevallier_Polarski_2001} for the 
Dark Energy equation of state parameter $w(z)$ and we include supernovae data
in order to better constrain the background. Furthermore we also analyze 
SKA-like data alongside a Euclid-like survey.

It is important to remark that in the forecasts for Euclid and SKA we combine galaxy clustering and lensing (shear) in a 
non-parametric way to predict the measurement errors on the RSD quantity $f\sigma_8(z)$ and on the expansion rate $E(z)\equiv H(z)/H_0$, 
as shown in \cite{Amendola:2013qna}. Lensing measures the lensing potential and  the expansion rate, while galaxy clustering measures the 
amplitude of the Newtonian potential, the RSD $f\sigma_8(z)$ and the expansion rate. The resulting Fisher matrix is marginalised over the 
lensing potential and over the spectrum amplitude to obtain the forecasted error on $f\sigma_8(z),E(z)$. These two quantities are model-independent 
in the sense that they do not depend on the power spectrum shape ($f\sigma_8(z) $ becomes model independent after marginalising over an overall constant, 
as explained later on), nor on the model of galaxy bias. We then use exclusively these two quantities to obtain constraints on the cosmological 
and modified gravity parameters. This ensures that our final constraints are model-independent in the sense explained above. The price to pay is 
that the constraints might become weaker than shown in other papers that assume either specific spectra shapes (e.g. $\Lambda $CDM), or bias models, 
or restrict in some way the modified gravity sector.

\section{Linear perturbations}

\label{sec:ii} We focus our attention on the evolution of linear
perturbations in the quasi-static limit (i.e. for scales significantly
inside the sound horizon, $c_{s}k/(aH)\gg1$), such that the terms
containing $k$ dominate over the time-derivative terms.\\

As shown in \cite{Amendola:2012ky}, for a general modified gravity theory,
one has the following equation for linear perturbation growth:

\begin{equation}
\delta''_{m}+(2+\frac{E'}{E})\delta'_{m}=\frac{3}{2}\Omega_{m}\delta_{m}Y(a,k),\label{delta}
\end{equation}
where $\delta_{m}$ is the matter density contrast, $\Omega_{m}=\Omega_{m0}a^{-3}/E^{2}$
is the matter density parameter and $E\equiv H/H_{0}$ is the dimensionless
Hubble function, which describes the background expansion. The function
$Y$ represents the effective gravitational constant for matter and
it is defined as 
\begin{equation}
Y(a,k)=-\frac{2k^{2}\Psi}{3(aH)^{2}\Omega_{m}\delta_{m}}\label{Y}
\end{equation}
As done previously \cite{Taddei:2014wqa}, we assume either
that baryons do not feel modified gravity (i.e., there is a suitable conformal or disformal factor in the metric coupled to baryons that cancel the effects of modified gravity) or that the local gravity
experiments occur in an environments where the extra force is negligible (screening mechanism).
In either case, we can escape most local gravity constraints. The time variation
of the product $GM$ is not screened, so one still has to confront with local constraints at the present time.
However, one might always design a "time-screened" model such that the present constraints on $\dot {(GM)}/(GM)$ vanish while being significant in the past,
so again local constraints can always be avoided. We will therefore neglect any local constraint in this paper. 

The background expansion is described by: 
\begin{equation}
E^{2}=\Omega_{m,0}(1+z)^{3}+(1-\Omega_{m,0})e^{3\int(1+w(z'))/(1+z')dz'}\label{eq:back}
\end{equation}
in which we consider the Chevallier-Polarski-Linder parametrization
\cite{Chevallier_Polarski_2001} for the dark energy equation of state:

\begin{equation}
w(z)=w_{0}+w_{a}\frac{z}{(1+z)}.\label{eq:eqstato}
\end{equation}
Therefore Eq. (\ref{delta}) can be written as

\begin{equation}
\delta''_{m}+(2+\frac{E'}{E})\delta'_{m}=\frac{3}{2}\frac{\delta_{m}}{a^{3}E^{2}}\Omega_{m,0}Y.\label{eq:a-1-1}
\end{equation}
As in \cite{Taddei:2014wqa}, we define the initial conditions parameter
$\alpha\equiv\delta'_{in}/\delta_{in}$. When we solve Eq. (\ref{eq:a-1-1}),
we choose the initial value at $z_{in}\approx3.5$, just beyond
the observed range. We assume for now $Y$ to be constant in space
and time. The equation is then $k$-independent and the result can
be directly compared to observations that are obtained in a finite
range of scales. The constant $\alpha$ can itself in principle be
scale dependent but for simplicity we assume this to not be the case.
Since observations have so far been performed in a relatively small
range of scales, this simplification is probably not very harmful
at least for as concerns current constraints.

In total we have  five parameters $\{\Omega_{m,0},w_{0},w_{a},Y,\alpha\}$,
where only the first three enter the background equation (Eq. (\ref{eq:back})),
plus an overall constant for growth data that we can marginalize over
analytically. We adopt uniform flat prior probabilities allowing for
fairly large ranges (several $\sigma$ larger than current constraints)
as we expect degeneracies between standard and modified gravity parameters
to widen the region of the parameter space with a non vanishing likelihood.
The parameter space is sampled both through a grid approach and with
an MCMC approach through the publicly available package \texttt{cosmomc}
\cite{Lewis:2002ah}, with a convergence diagnostic using the Gelman
and Rubin statistic. We have thoroughly checked that the two approaches
give compatible results and we will use those obtained with the latter
throughout this paper.

Later on, we will take the Horndeski space-dependent form and we will
have to define the $k$-ranges. In this case the expression for $Y$
in the quasi-static limit can be written as \cite{Amendola:2012ky}

\begin{equation}
Y=h_{1}\frac{1+(k/k_{p})^{2}h_{5}}{1+(k/k_{p})^{2}h_{3}}\label{Y1-1}
\end{equation}
where $h_{1},h_{3},h_{5}$ are time dependent functions that can be
explicitly obtained when the full Horndeski Lagrangian is given \cite{2013PhRvD..87b3501A}.
The scale $k_{p}$ is an arbitrary pivot scale that we choose to be
$k_{p}=1h$/Mpc.

\section{Likelihood analysis}

\label{sec:iii}

We describe the growth rate data as a set of values ${f\sigma_{8}}_i$
at various redshifts, defined as:

\begin{equation}
{f\sigma_{8}}_i=f(z_{i})\sigma_{8}(z_{i})=f(z_{i})\sigma_{8}G(z_{i})=\sigma_{8}\frac{\delta_{i}'}{\delta_{0}}=\sigma_{8}d_{i}\label{eq:growth}
\end{equation}
where $d_{i}=\delta_{i}^{'}/\delta_{0}$. The growth-rate theoretical
estimates are defined as $\hat{d}_{i}=\hat{\delta}'_{i}/\hat{\delta}_{0}$,
where $\hat{\delta}$ is the solution of Eq. (\ref{eq:a-1-1}).
Also for the background, we can define the data $E_{i}$
and the theoretical estimates $\hat{E}_{i}$, given
by Eq. (\ref{eq:back}), calculated for every redshift values. We
build then the $\bar{\chi}_{}^{2}$ function:
\begin{equation}
\bar{\chi}_{}^{2}=(\vec{d}_{i}-\vec{t}_{i})C_{ij}^{-1}(\vec{d}_{j}-\vec{t}_{j})\label{eq:chifull}
\end{equation}
in which the data vector $\vec{d}_{i}$ contains $\{{f\sigma_{8}}_i;E_{i}\}$,
$\vec{t}_{i}$ includes $\{\hat{d}_{i};\hat{E}_{i}\}$ for each
value of redshift $z_{i}$ and $C_{ij}$ is the covariance matrix
of the data. We decompose $C_{ij}^{-1}$ into three sub-matrices $(RR),(ER),(EE)$
as

\begin{equation}
C_{ij}^{-1}=\left(\begin{array}{cc}
(RR)_{ij} & (ER)_{ij}\\
(ER)_{ij} & (EE)_{ij}
\end{array}\right)\label{eq:covfull}
\end{equation}
They are respectively the growth-rate error matrix $(RR)_{ij}$, the
background error matrix $(EE)_{ij}$ and the mixing matrix between
the growth-rate and the background errors $(ER)_{ij}$.

As mentioned in the Introduction, the data $f\sigma_{8}(z)$ are defined
up to a constant that depends on the choice of the shape of the power
spectrum. Here we wish to remain as much model-independent as possible
and therefore we marginalize over this overall constant. Once again,
in principle the constant might be scale dependent but we simplify
our task by assuming it is not. Our goal, after all, is to see how
much the constraints change when some of the model-dependent assumptions
are removed and it is sufficient for now to show the effect when just
a minimal set of assumptions are removed.

We marginalize then the likelihood $L'=\exp(-\bar{\chi}^{2}/2)$ over
an overall positive factor with a uniform prior; this leads to a new
posterior $L=\exp(-\chi{}^{2}/2)$ where 
\begin{equation}
\chi{}^{2}=S_{tt}-\frac{S_{dt}^{2}}{S_{dd}}-2\log\left(1+\mathrm{Erf}(\frac{S_{dt}}{\sqrt{2S_{dd}}})\right)\label{eq:chimarg}
\end{equation}
and: 
\begin{eqnarray}
S_{dt} & = & d_{i}(RR)_{ij}\hat{d}_{ j}+\left(\hat{E}_{i}-E_{i}\right)(ER)_{ij}d_{ j}\\
S_{dd} & = & d_{i}(RR)_{ij}d_{ j}\\
S_{tt} & = & \left(E_{i}-\hat{E}_{i}\right)(EE)_{ij}\left(E_{j}-\hat{E}_{j}\right)-2(E_{i}-\hat{E}_{i})(ER)_{ij}\hat{d}_{ j}+\hat{d}_{i}(RR)_{ij}\hat{d}_{ j}\label{eq:Stt}
\end{eqnarray}
This is the likelihood we will employ when forecasting constraints from future experiments.

When estimating the forecast constrains from clustering and lensing, we assume $E(z)$ as a free parameter in each bin, and therefore the final constraints on $E(z)$ will be correlated with thos on the clustering and lensing variables. Current data, however, are produced in such a way that
$E(z)$ either does not enter the analysis or is fixed to $\Lambda$CDM. In both cases, it is not correlated with the growth data.
The procedure for the current data is therefore slightly different  and will be discussed
next.

\section{Current data}

\label{sec:iv}

\subsection*{Supernovae data}

In order to constrain the background, we will use the most recent
SN catalog dubbed JLA (acronym for Joint Lightcurve Analysis \cite{Betoule:2014frx}).
This is a joint analysis of the 740 spectroscopically confirmed supernovae
type Ia of the SNLS and SDSS-II collaborations and covers a redshift
range from $0.02$ to $1.3$. We can define the predicted magnitudes
$m(z_{i})_{th}$ by: 
\begin{equation}
m(z_{i})_{th}=M+5\log_{10}d_{L}(z_{i})+25\label{eq:mu-1}
\end{equation}
where $d_{L}(z)$ is the luminosity distance, which is computed under
the assumption of spatial flatness:

\begin{equation}
d_{L}(z)=\frac{(1+z)}{H_{0}}\intop_{0}^{z}\frac{dz'}{E(z')}\label{eq:dislum-1}
\end{equation}
and $M$ is the absolute magnitude. We can rewrite Eq. (\ref{eq:mu-1})
as $m(z_{i})_{th}=\mu(z_{i})_{th}+\gamma$, where $\mu(z_{i})_{th}=5\log_{10}\hat{d}_{L}(z_{i})$
(with $\hat{d}_{L}\equiv d_{L}H_{0}$) and $\gamma=M+25-5\log_{10}H_{0}$.
We note that, instead of the background theoretical estimates of $E_{i}$,
we have the theoretical estimates of the apparent magnitude, defined
in Eq. (\ref{eq:mu-1}).

The observed apparent magnitude $m_{i}$ is given by \cite{Betoule:2014frx}:
\begin{equation}
m=m_{B}^{*}-M_{B}+\alpha X_{1}-\beta C\label{eq:appmagn-1-1}
\end{equation}
where $m_{B}^{*}$ corresponds to the observed peak magnitude in the
rest-frame $B$ band, $X_{1}$ and $C$ describe, respectively, the
time stretching of the light-curve and the supernova color at maximum
brightness, and $\alpha,\beta,M_{B}$ are nuisance parameters related
to the distance estimate, which are added to the sampled parameters
and then marginalized out. % The absolute magnitude $M_{B}$% is related to the host stellar mass $(M_{stellar})$ by the simply% step function, which is described in \cite{Betoule:2014frx}. We have% fixed $\alpha,\beta,M_{B}$ to the best fit for $\Lambda$CDM model (see % \cite{Betoule:2014frx}).Thelikelihood
The analysis for the supernovae is based on the $\chi'^{2}$ function:
\begin{equation}
\chi_{SNIa}^{'2}=\sum_{i}\frac{[m_{i}-\mu(z_{i})_{th}-\gamma]^{2}}{\sigma_{i}^{2}}\label{eq:chi2sn-1}
\end{equation}
where $\sigma_{i}$ are the errors on the apparent magnitudes and
the index $i$ labels the elements of the JLA dataset. We marginalize
the likelihood $L'_{SNIa}=\exp(-\chi_{SNIa}^{'2}/2)$ over $\gamma$
\cite{Amendola2010}, leading to a new marginalized $\chi^{2}$ function:
\begin{equation}
\chi_{SNIa}^{2}=\left(S_{2}-\frac{S_{1}^{2}}{S_{0}}\right)\label{eq:chiSnmarg-1}
\end{equation}
where the quantities $S_{n}$ are defined as: 
\begin{equation}
S_{n}\equiv\sum_{i}\frac{[m{}_{i}-\mu(z_{i})_{th}]^{n}}{\sigma_{i}^{2}}.
\end{equation}

\subsection*{Growth data}

The current dataset includes all the independent published estimates
of $f\sigma_{8}(z)$ obtained with the redshift distortion method.
It includes the data from 2dFGS, 6dFGS, LRG, BOSS, CMASS, WiggleZ
and VIPERS, and spans the redshift interval from $z$ = 0.07 to $z$
= 0.8, see Table \ref{tab:Current-published-values} (see also \cite{Macaulay:2013swa,More:2014uva}).
In some cases the correlation coefficient between two samples has
been estimated in \cite{Macaulay:2013swa} and included in our analysis;
when there are different published results from the same dataset in
Table \ref{tab:Current-published-values}, we include only the more
recent one.

The $\chi_{growth}^{2}$ can be expressed as in Eq.(\ref{eq:chimarg})
with coefficients 
\begin{eqnarray}
S_{dt} & = & d_{i}(RR)_{ij}\hat{d}_{ j}\\
S_{dd} & = & d_{i}(RR)_{ij}d_{ j}\\
S_{tt} & = & \hat{d}_{i}(RR)_{ij}\hat{d}_{j}\label{eq:SttSn}
\end{eqnarray}
We can then combine these data with the marginalized SN $\chi^{2}$
of Eq.(\ref{eq:chi2sn-1}) to include information on the background
\begin{equation}
\chi{}^{2}=\chi_{growth}^{2}+\chi_{SNIa}^{2}\label{eq:chitot}
\end{equation}

\begin{center}
\begin{table}
\centering{}%
\begin{tabular}{|c|c|c|c|}
\hline 
Survey  & z  & $f(z)$ $\sigma_{8}(z)$  & References\tabularnewline
\hline 
\hline 
6dFGRS  & 0.067  & 0.423 $\pm$ 0.055  & Beutler et al. (2012) \cite{Beutler:2012px}\tabularnewline
\hline 
\multirow{2}{*}{LRG-200}  & 0.25  & 0.3512 $\pm$ 0.0583  & \multirow{2}{*}{Samushia et al (2012) \cite{Samushia:2011cs}}\tabularnewline
%\hline 
\cline{2-3}  & 0.37  & 0.4602 $\pm$ 0.0378  & \tabularnewline
\hline 
\multirow{2}{*}{LRG-60}  & 0.25{*}  & 0.3665$\pm$0.0601  & \multirow{2}{*}{Samushia et al (2012) \cite{Samushia:2011cs}}\tabularnewline
%\hline 
\cline{2-3} & 0.37{*}  & 0.4031$\pm$0.0586  & \tabularnewline
\hline 
\multirow{2}{*}{BOSS}  & 1) 0.30  & 0.408$\pm$ 0.0552, $\rho_{12}=-0.19$  & \multirow{2}{*}{Tojeiro et al. (2012)\cite{Tojeiro:2012rp}}\tabularnewline
%\hline 
\cline{2-3} & 2) 0.60  & 0.433$\pm$ 0.0662  & \tabularnewline
\hline 
\multirow{3}{*}{WiggleZ}  & 1) 0.44  & 0.413 $\pm$ 0.080, $\rho_{12}=0.51$  & \multirow{3}{*}{Blake (2011) \cite{Blake:2012pj}}\tabularnewline
%\hline 
\cline{2-3} & 2) 0.60  & 0.390 $\pm$ 0.063, $\rho_{23}=0.56$  & \tabularnewline
%\hline 
\cline{2-3} & 3) 0.73  & 0.437 $\pm$ 0.072  & \tabularnewline
\hline 
Vipers  & 0.8  & 0.47 $\pm$ 0.08  & De la Torre et al (2013)\cite{delaTorre:2013rpa}\tabularnewline
\hline 
2dFGRS  & 0.17  & 0.51 $\pm$ 0.06  & Percival et al. (2004) \cite{Percival:2004fs}\cite{Song:2008qt}\tabularnewline
\hline 
LRG  & 0.35  & 0.429 $\pm$ 0.089  & Chuang and Wang (2013) \cite{Chuang:2012qt}\tabularnewline
\hline 
LOWZ  & 0.32  & 0.384$\pm$0.095  & \multirow{1}{*}{Chuang at al (2013)\cite{Chuang:2013wga}}\tabularnewline
\hline 
\multirow{4}{*}{CMASS}  & 0.57{*}  & 0.348 $\pm$ 0.071  & \multirow{1}{*}{}\tabularnewline
%\hline 
\cline{2-4} & 0.57{*}  & 0.423 $\pm$ 0.052  & Beutler et al (2014)\cite{Beutler:2013yhm}\tabularnewline
%\hline 
\cline{2-4} & 0.57  & 0.441$\pm$0.043  & Samushia et al (2014) \cite{Samushia:2013yga}\tabularnewline
%\hline 
\cline{2-4} & 0.57{*}  & 0.450 $\pm$ 0.011  & Reid et al (2013)\cite{Reid:2014iaa}\tabularnewline
\hline 
\end{tabular}\protect\protect\protect\caption{\label{tab:Current-published-values}Current published values of $f\sigma_{8}(z)$.
In some cases we list also the correlation coefficient $\rho_{ij}$
between different bins \cite{Macaulay:2013swa}. When there are different
published results from the same dataset, we include only the more
recent one (we put an asterisk to the entries which are not employed
in this analysis).}
\end{table}

\par\end{center}

\subsection*{Results}

We now analyze current data using the prescription of Eq.(\ref{eq:chitot})
in two different cases: the $\Lambda$CDM case, in which we fix $Y=1$
and $\alpha=1$, and the case where the additional parameters are
free to vary, from now on denoted ``MG case''.

\begin{figure}[t]
\begin{center}
\includegraphics[bb=0bp 0bp 236bp 242bp,width=7cm]{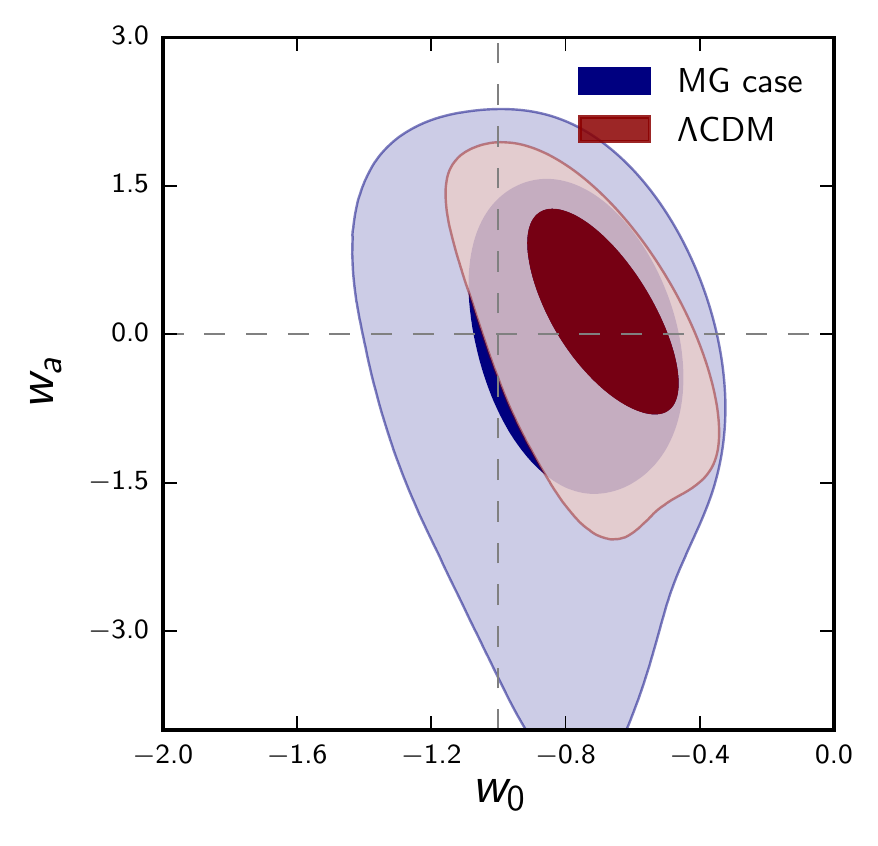}
\includegraphics[bb=0bp 0bp 239bp 242bp,width=7cm]{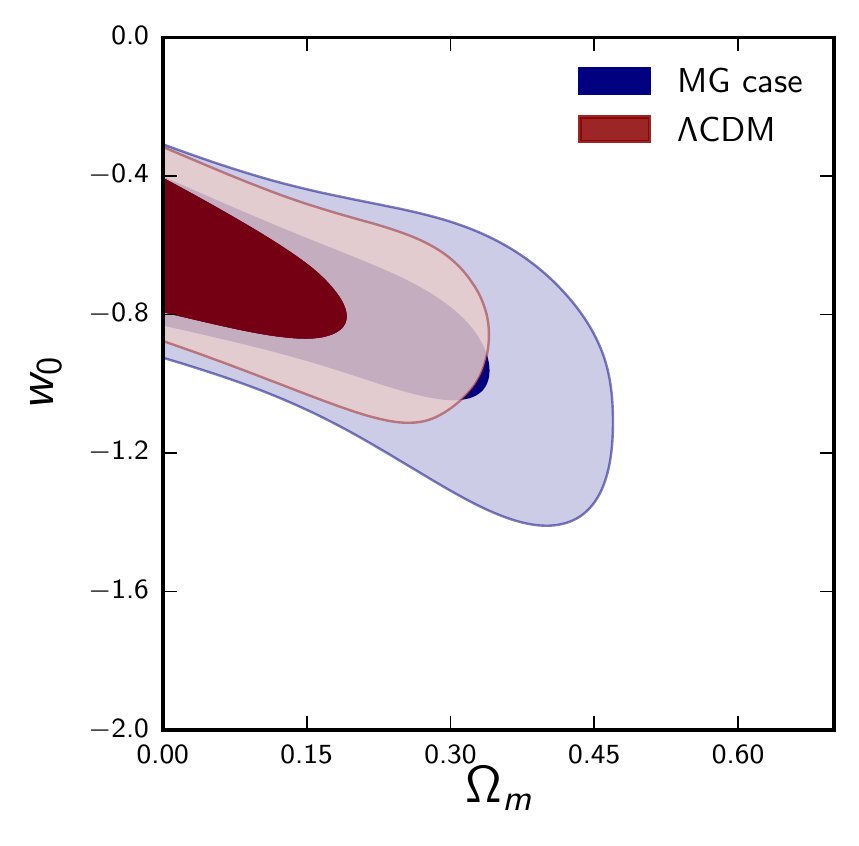}\\
 \includegraphics[bb=0bp 0bp 236bp 242bp,width=7cm]{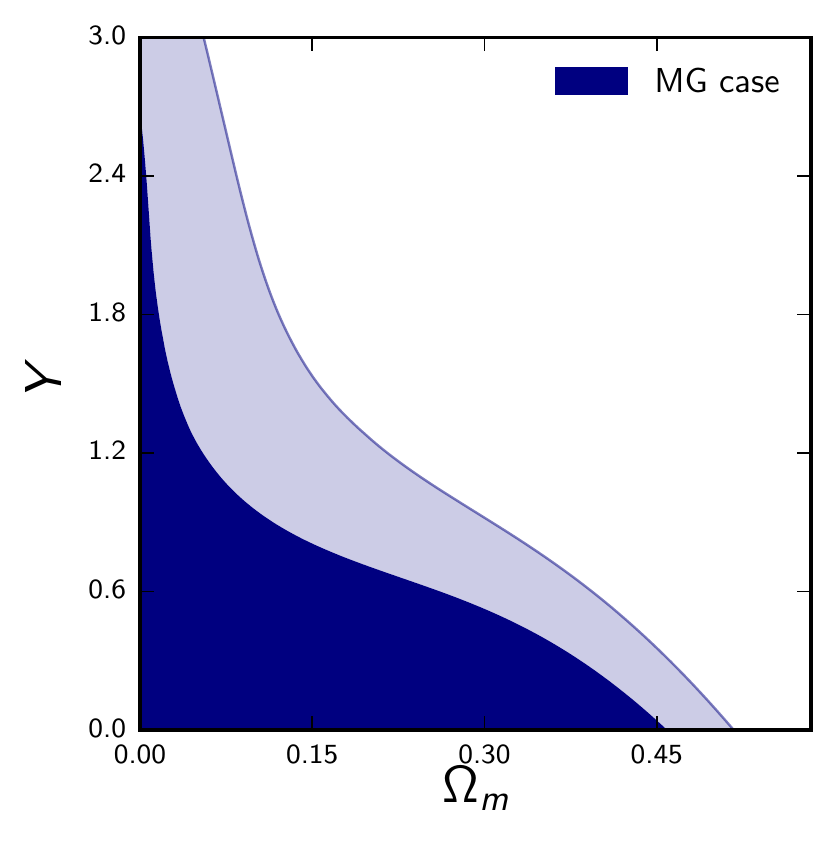}
\includegraphics[bb=0bp 0bp 239bp 242bp,width=7cm]{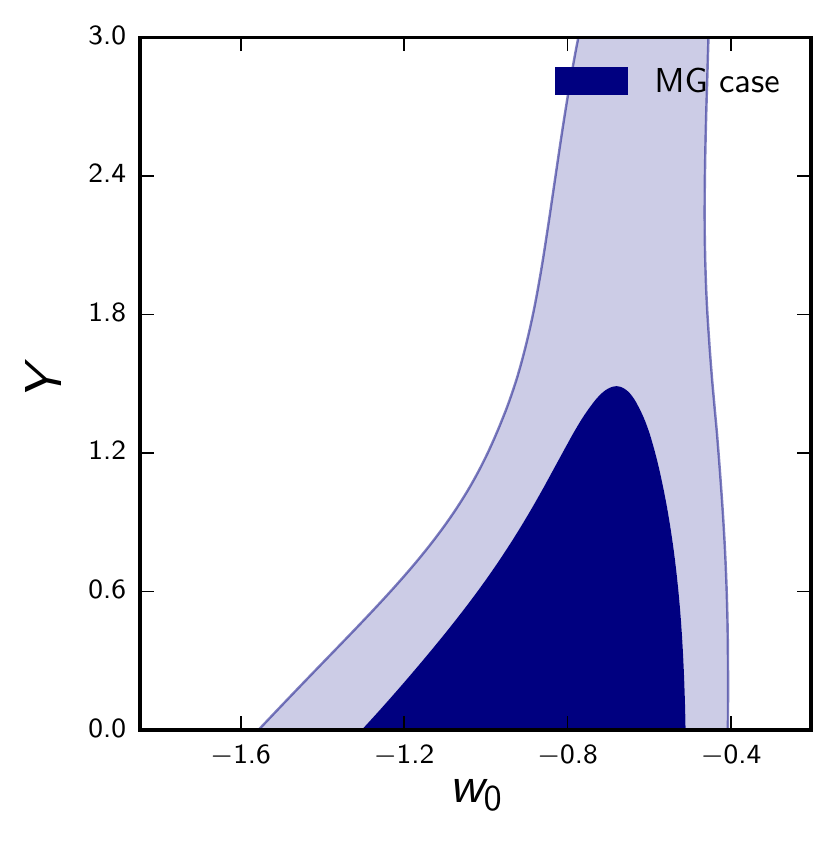}
\protect\protect\protect\caption{Top panels: 1$\sigma$ and 2$\sigma$ confidence-level contours for
the parameters $\{w_{0},w_{a}\}$ (top left), $\{\Omega_{m,0},w_{0}\}$
(top right). Bottom panels: confidence-level contours for the parameters
$\{\Omega_{m,0},Y\}$ (bottom left) and $\{w_{0},Y\}$ (bottom right).
Red contours refer to the $\Lambda$CDM analysis, while blue ones
refer to the full case.}
\label{fig:curres} 
\end{center}
\end{figure}

Table \ref{tab:currentres} and Fig. \ref{fig:curres} show the results
obtained for these two analysis; the top panels of Fig. \ref{fig:curres}
highlight how the inclusions of $Y$ and $\alpha$ in the analysis
bring to looser constraints on standard cosmological parameters 
due the degeneracies highlighted in the bottom panel. In particular,
with respect to the $\Lambda$CDM case,
the 2-$\sigma$ bound on $\Omega_m$ increases by $41\%$, while the 
1-$\sigma$ error on $w_0$ and $w_a$ increases by $44\%$ and $53\%$ respectively.
% In particular, at 95\% 
% c.l., the uncertainty on $\Omega_{m,0}$ increases of 0.07, while the uncertainties on $w_{0}$
% and $w_{a}$ increase of 0.04 and 0.3 respectively at 68\% c.l., when the full 
% case is considered.
Furthermore, it is interesting to notice how the inclusion of MG parameters
relaxes the slight tension of current data with the $\Lambda$CDM
background ($w_{0}=-1,\ w_{a}=0$) at the cost of non standard best fit values
for $Y$ and $\alpha$.

\begin{figure}[!h]
\begin{centering}
\hspace*{-1cm} %
\begin{tabular}{cc}
\includegraphics[width=7cm]{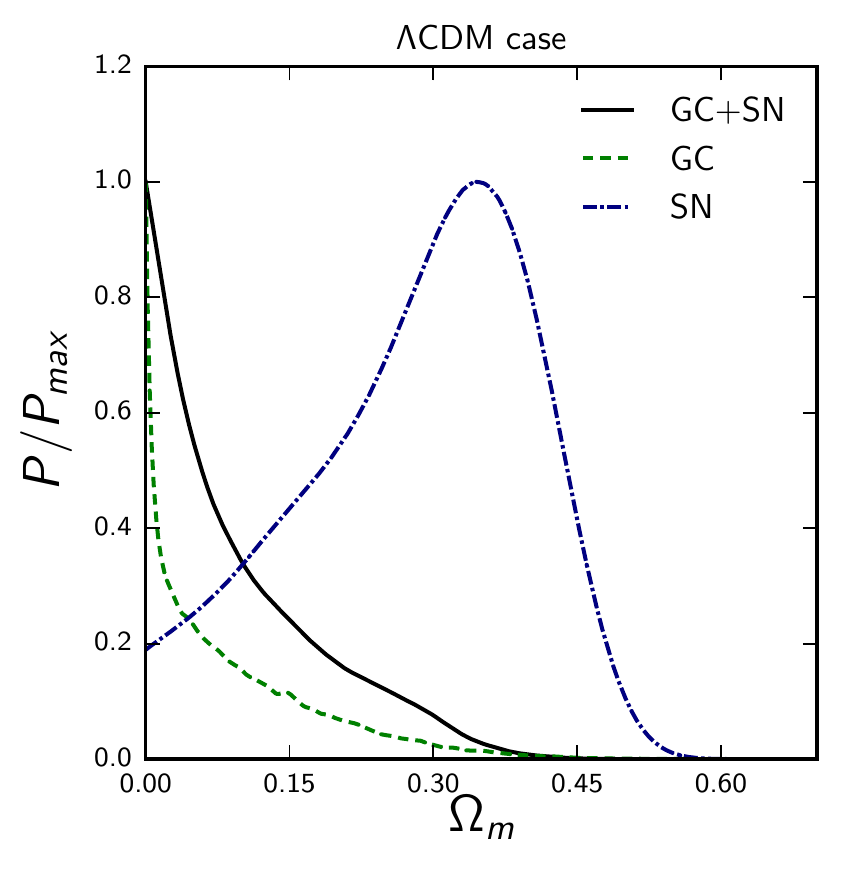}  & \includegraphics[width=7cm]{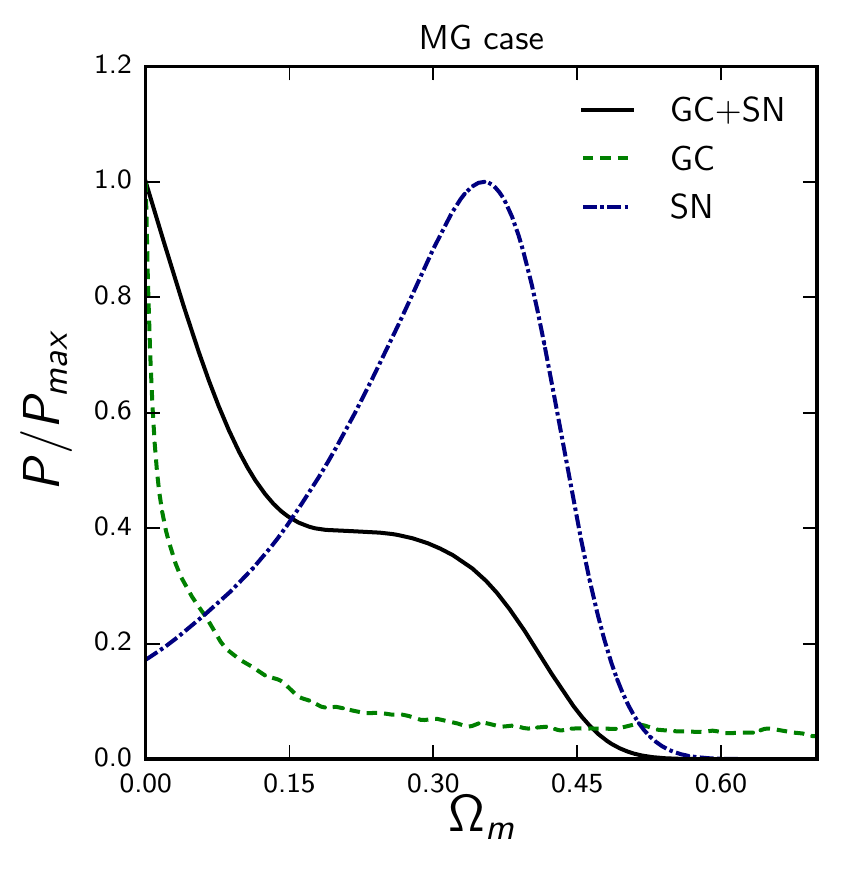} \tabularnewline
\end{tabular}\protect\protect\caption{Posterior distribution of $\Omega_{m}$ in the $\Lambda$CDM (left
panel) and MG (right panel) cases, using growth data (green dashed
line), supernovae data (blue dot-dashed line) and their combination
(black solid line).}
\par
\label{fig:omobs} 
\end{centering}
\end{figure}

Fig. \ref{fig:omobs} shows the posterior distribution obtained on
$\Omega_{m,0}$ through growth data, supernovae and their combination
in the MG case; it is interesting to notice that the growth rate data
tend to move the marginalized 1-dimensional posterior for $\Omega_{m,0}$
toward smaller values, while supernovae data support a maximum near
$\Omega_{m,0}=0.3$. The combined analysis still prefers low values
of the matter density, but the effect of supernovae is to make more
likely also higher values with respect to the growth-data-only analysis.
This effect is stronger in the MG case as the degeneracies of $Y$
and $\alpha$ with $\Omega_{m,0}$ degrade the constraining power
of growth data on the matter density, increasing therefore the statistical
significance of the SN dataset.

\begin{figure}[!h]
\begin{center}
\includegraphics[bb=0bp 0bp 236bp 242bp,width=7cm]{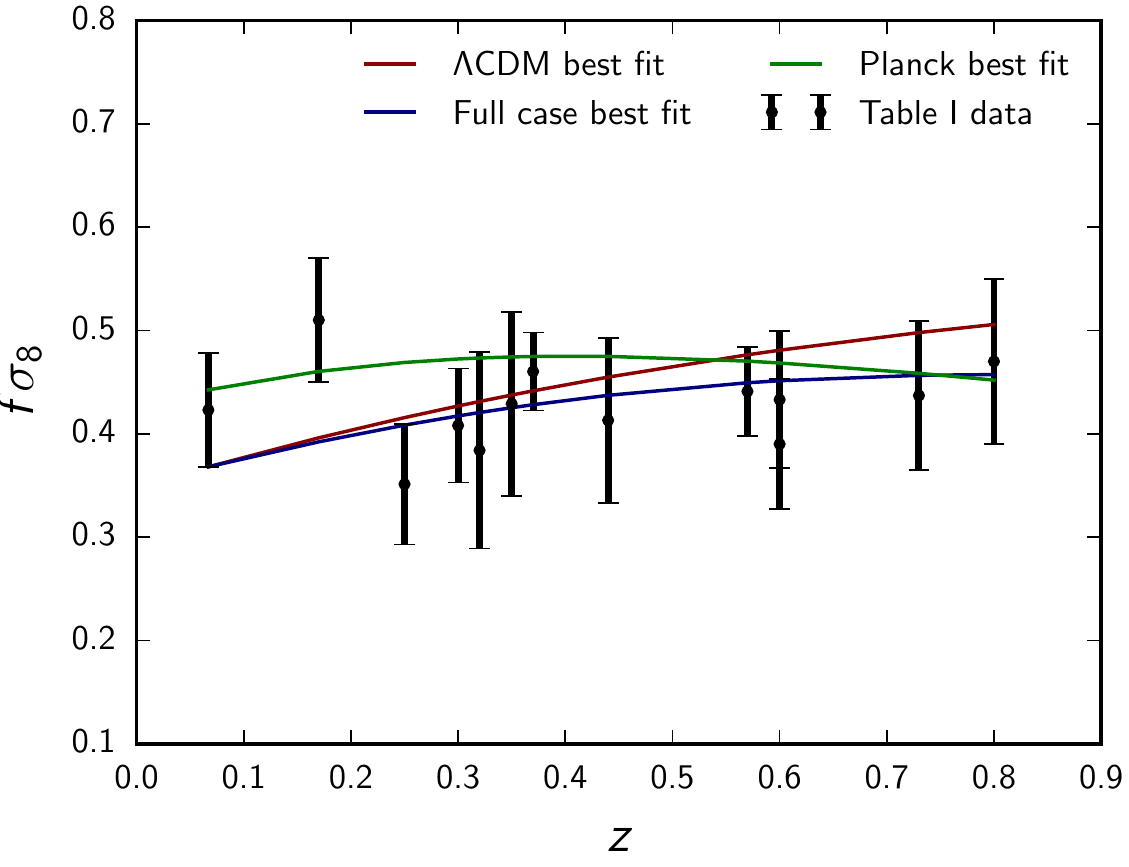}
\protect\protect\protect\caption{Currently available $f\sigma_{8}(z)$ data compared with the theoretical
prediction using Planck 2015 best fit parameters (green line) and
with the $\Lambda$CDM and MG cases best fits (respectively red and
blue lines).}

\label{fig:bfplot} 
\end{center}
\end{figure}

Fig. \ref{fig:bfplot} shows the best fit $f\sigma_8(z)$ trends found with
our analysis (both in the $\Lambda$CDM and MG cases) compared to
the $f\sigma_{8}(z)$ predicted using the best fit results of Planck
2015 \cite{Ade:2015xua}. Our marginalization over $\sigma_{8}$ allows
to rescale the theoretical $f\sigma_8(z)$ by an arbitrary factor and this
allows non standard values of $\Omega_{m,0}$ and $w_{0},\ w_{a}$ to
better fit the data; comparing the obtained best fit $\chi^2$ values with the 
one given by the Planck best fit parameters, we find $\Delta\chi^2=-2.57$ in 
the $\Lambda$CDM case, while allowing also for $Y$ and $\alpha$ to vary we obtain
an improvement $\Delta\chi^2=-3.25$.

\begin{table}
\begin{center}
\begin{tabular}{|c|c|c|c|c|c|}
\hline 
\multirow{1}{*}{\textbf{Current Data}}  & \multirow{1}{*}{$\Omega_{m,0}$ }  & \multirow{1}{*}{$w_{0}$}  & \multirow{1}{*}{$w_{a}$}  & \multirow{1}{*}{Y}  & \multirow{1}{*}{$\alpha$} \tabularnewline
\hline 
$\Lambda$CDM case  & $<0.273$  & $-0.71_{-0.11}^{+0.16}$  & $0.17_{-0.53}^{+0.71}$  & fixed to 1  & fixed to 1\tabularnewline
\hline 
MG case  & $<0.409$  & $-0.79_{-0.11}^{+0.24}$  & $-0.29_{-0.52}^{+1.4}$  & $<2.22$  & $0.21_{-0.72}^{+0.21}$ \tabularnewline
\hline 
\end{tabular}

\protect

\protect\protect\protect\protect\caption{Marginalized values and 1-$\sigma$ errors on the free parameters
for current data in the $\Lambda$CDM case and in the MG case. When
a two-tail likelihood is not allowed by the data, the 2-$\sigma$
upper limit is shown.}

\label{tab:currentres} 
\end{center}
\end{table}

% 
% \begin{table}
% \begin{tabular}{|c|c|}
% \hline 
%  & $\Delta\chi^{2}$ \tabularnewline
% \hline 
% $\Lambda$CDM case  & $-2.57$ \tabularnewline
% \hline 
% MG case  & $-3.25$ \tabularnewline
% \hline 
% \end{tabular}
% 
% \protect
% 
% \protect\protect\protect\protect\caption{$\Delta\chi^{2}$ for the $\Lambda$CDM and MG cases with respect
% to Planck 2015 best fit. \textbf{(do we need this? I think is better
% in the text)} }
% 
% 
% \label{tab:deltachi} 
% \end{table}

\section{Forecast data}

\label{sec:v}

\subsection*{Euclid mission}

After having obtained constraints from currently available data, it
is interesting to investigate how future surveys can improve our tests
of gravity. To this purpose we focus on the future ESA mission Euclid,
using the specifications presented in \cite{Euclid-r}, 
i.e. a coverage of 15,000 square degrees and a redshift
interval $z=0.5-1.5$. In order to forecast $f\sigma_{8}$ data from
this survey, we follow the $z$-binning procedure of \cite{Amendola:2013qna}
and we divide the redshift range in equally spaced bins of width $\Delta z$
= 0.2 and, in order to prevent accidental degeneracy due to low statistic,
a single larger redshift bin between $z=1.5-2.1$, for a total of
six bins. As we are mainly interested in the 
effect of the theory approach generality on constraints, we limit our analysis
to the nominal Euclid specifications \cite{Euclid-r}; however, recently 
the modelling of $H\alpha$ emitters luminosity function has been discussed \cite{Pozzetti:2016cch}
and we emphasize that also this more conservative approach in forecasting the amount of observed
sources might have a significant effect on the expect results.

Beside this, we also include in our
forecast the Euclid weak lensing forecasts, 
following the prescription of \cite{Amendola:2013qna}, where
the full Fisher matrix for redshift distortion, galaxy clustering and
weak lensing has been estimated.

The fiducial values and errors for $f\sigma_{8}$ and $E$ are summarized
in Table \ref{euclidall} (see also Fig. \ref{fig:errors}). As with
currently available data, we analyze the $\Lambda$CDM and the MG
cases, investigating how the additional parameters affect the achievable
constraints. We use now the $\chi^{2}$ described in Eq. (\ref{eq:chimarg}),
sampling the parameter space with the techniques mentioned at the
end of  Section 2.% \ref{sec:ii}.

As done for the currently available data, we assume that the additional
parameters $Y$ and $\alpha$, together with the galaxy bias $b$, are not 
scale dependent; for this reason we consider only the redshift binning, without 
including any binning in $k$. An explicit scale dependence for $Y$ parameter is 
instead considered in Section \ref{sec:vi} of this paper, where therefore the 
predicted observations performed by the Euclid survey are binned also in scale.

\begin{table}[htbp]
\begin{centering}
\begin{tabular}{|l|c|c|c|c|}
\hline 
\multicolumn{3}{|c|}{Fiducial} & \multicolumn{2}{c|}{Euclid} \tabularnewline
\hline 
$\bar{z}$  & $E$  & $f$  & $\Delta E$ (68\% c.l.)  & $\Delta f\sigma_{8}$ (68\% c.l.)  \tabularnewline
\hline 
$0.6$  & $1.37$  & $0.469$  & $0.011$  & $0.0078$  \tabularnewline
\hline 
$0.8$  & $1.53$  & $0.457$  & $0.021$  & $0.0057$  \tabularnewline
\hline 
$1.0$  & $1.72$  & $0.438$  & $0.024$  & $0.0045$  \tabularnewline
\hline 
$1.2$  & $1.92$  & $0.417$  & $0.024$  & $0.0038$  \tabularnewline
\hline 
$1.4$  & $2.14$  & $0.396$  & $0.025$  & $0.0035$  \tabularnewline
\hline 
$1.8$  & $2.62$  & $0.354$  & $0.034$  & $0.0025$  \tabularnewline
\hline 
\end{tabular}
\par\end{centering}

\protect\protect\protect\caption{ Fiducial values and errors for $f\sigma_8$ and $E$
for the Euclid survey, combining weak lensing and galaxy clustering measurements (from \cite{Amendola:2013qna}).}

\label{euclidall} 
\end{table}

\subsection*{SKA mission}

As a comparison with Euclid, we also consider another almost contemporary
facility, the Square Kilometre Array (SKA), which is a giant radio
telescope array which will be built across two sites, in South Africa
and in Western Australia. Its main aim will be to probe the nature
of dark energy by mapping out large-scale structures, primarily using
the 21-cm emission line of neutral hydrogen (HI) to measure the galaxy's
redshift with high precision \cite{Abdalla:2015zra}.

The SKA project will consist of two phases, denoted SKA1 and SKA2 in the following. Phase 1 will be formed
by three different arrays: SKA1-MID, SKA1-SUR, which are dish arrays
with a mid-frequency ($\nu\apprle$ 1.4 GHz) receivers and low/intermediate
redshifts and SKA1-LOW which is an aperture array with a lower frequency
($\nu<$ 350 MHz) and higher redshifts. The LOW will be capable to
detect HI emission only for $z\geq3$ by using intensity mapping (IM)
rather then a galaxy survey, so we will not consider it here. The
SKA1 survey (for either MID and SUR) will cover 5,000 deg$^{2}$ area
and it will detect about 5$\times$10$^{6}$ HI galaxies up to $z\sim0.5$. The second phase,
 SKA2, planned for the late 2020s,  will have a sensitivity  around 10 times
the one of SKA1 and should be capable of detecting about 10$^{9}$ HI
galaxies over a 30,000 deg$^{2}$ area, up to $z\sim2.0$. 

 For both cases, we follow the procedure described in \cite{Yahya:2014yva}
and obtain the galaxy redshift distribution functions ($dN/dz$) and
the bias redshift evolution ($b(z)$) as 
\begin{eqnarray}
\frac{dN}{dz}(z) & = & 10^{c_{1}}z^{c_{2}}\exp{(-c_{3}z)}\\
b(z) & = & c_{4}\exp{(c_{5}z)}
\end{eqnarray}
where the values of the $c_{i}$ parameters are obtained from \cite{Yahya:2014yva}.

\begin{figure}[!h]
\begin{center}
\hspace*{-1cm} %
\begin{tabular}{cc}
\includegraphics[width=8cm]{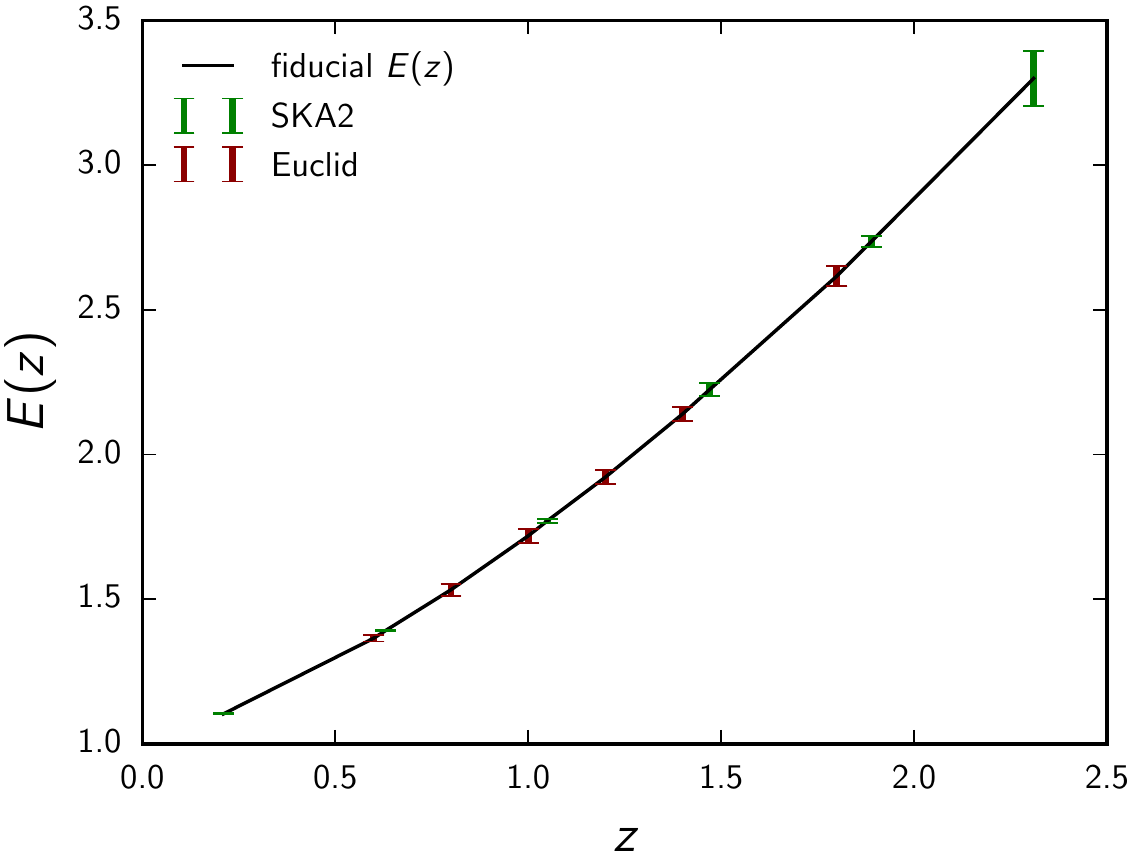}  & \includegraphics[width=8cm]{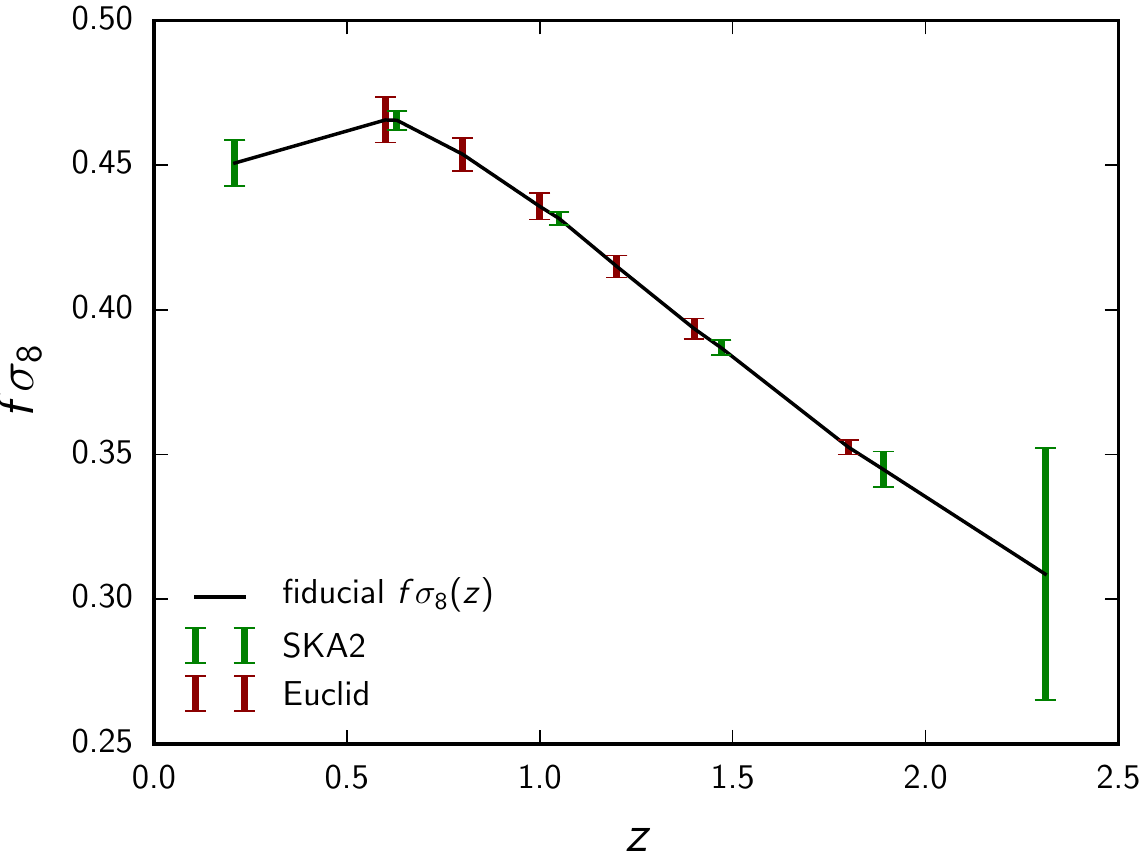} \tabularnewline
\end{tabular}\protect\protect\caption{Fiducial cosmology predictions for $E(z)$ and $f\sigma_{8}(z)$ (black
lines) and forecast errors obtained with specifications for SKA2 (green error bars) and Euclid Full (red error bars).}
\label{fig:errors} 
\par\end{center}

\end{figure}

In \cite{Yahya:2014yva}, constraints on dark energy equation of state
parameters $w_{0}$ and $w_{a}$ and on the spatial curvature parameter
$\Omega_{K}$ are obtained using forecast BAO measurements for SKA1
and SKA2 and these are compared with a Euclid like H$\alpha$ galaxy
survey. The interesting point is that SKA2 outperforms Euclid by a
factor around 2 due to the fact that it has a double area and an additional
redshift bin below Euclid's minimum redshift. As for Euclid, we build
forecast datasets for both SKA1 and SKA2 observations of galaxy clustering,
dividing the redshift range $0\leq z\leq2.5$ in six equispaced bins,
which are reported in Table \ref{skaall} (see also Fig. \ref{fig:errors}),
and we analyze these in the $\Lambda$CDM and MG cases. As done for Euclid 
observations, we do not consider a $k$-binning for this case.

SKA  will also be able to perform shear measurements as the Euclid survey \cite{Harrison:2016stv,Bonaldi:2016lbd,Camera:2016owj}.
Therefore, following the same procedure used for Euclid and the SKA shear specifications \cite{Harrison:2016stv}, we include
forecast shear measurements  for both SKA1 and SKA2.

\begin{table}[!h]
\begin{centering}
\begin{tabular}{|l|c|c|c|c|c|c|}
\hline 
\multicolumn{3}{|c|}{Fiducial} & \multicolumn{2}{c|}{SKA1} & \multicolumn{2}{c|}{SKA2}\tabularnewline
\hline 
$\bar{z}$  & $E$  & $f\sigma_{8}$  & $\Delta E$ (68\% c.l.)  & $\Delta f\sigma_{8}$ (68\% c.l.)  & $\Delta E$ (68\% c.l.)  & $\Delta f\sigma_{8}$ (68\% c.l.)\tabularnewline
\hline 
$0.21$  & $1.10$  & $0.455$  & $0.03$  & $0.022$  & $0.002$  & $0.008$\tabularnewline
\hline 
$0.63$  & $1.39$  & $0.468$  & $0.02$  & $0.019$  & $0.001$  & $0.003$\tabularnewline
\hline 
$1.05$  & $1.77$  & $0.433$  & $0.12$  & $0.5$  & $0.007$  & $0.002$\tabularnewline
\hline 
$1.47$  & $2.22$  & $0.388$  & $0.16$  & $39$  & $0.02$  & $0.003$\tabularnewline
\hline 
$1.89$  & $2.73$  & $0.346$  & $0.69$  & $3.4\ 10^3$  & $0.02$  & $0.006$\tabularnewline
\hline 
$2.31$  & $3.29$  & $0.309$  & $1.00$  & $3.0\ 10^5$  & $0.09$  & $0.04$\tabularnewline
\hline 
\end{tabular}
\par\end{centering}

\protect\protect\protect\protect\protect\protect\protect\protect\protect\caption{Fiducial values and SKA1 and SKA2 errors for $f\sigma_{8}$ and
$E$ combining weak lensing and galaxy clustering measurements.}

\label{skaall} 
\end{table}

\subsection*{Future Supernovae survey}
On top of the information brought by Weak Lensing and Galaxy Clustering,
we assume that constraints on the background are also imposed by a
future supernovae Ia survey; we consider observations in the
redshift range 0 $<z<$ 2.5 and we divide the interval in the same bins
used for the SKA survey (but with the initial redshift $z=0$ instead of $0.05$).
We also assume that, in this range, the total number of observed SN
is  $n_{SN} = 10^5$ as expected for LSST survey \cite{Tyson:2003kb}, taking as a reference
the Union 2.1 \cite{Amanullah:2010vv} catalogue for the distribution
of data in each bin and for the average magnitude errors. Notice that with this choice the last two SKA bins do not include any supernova.

Given these survey specifications we obtain the Fisher matrix for supernovae observation and the $\chi^2$ 
obtained from these at each step of the MCMC is then added to the one obtained by the combination WL+GC in order to construct
a joint posterior for the free parameters.

\subsection*{Results}

\begin{figure}[!h]
\begin{center}
\begin{tabular}{cc}
\includegraphics[bb=0bp 0bp 236bp 242bp,width=6cm]{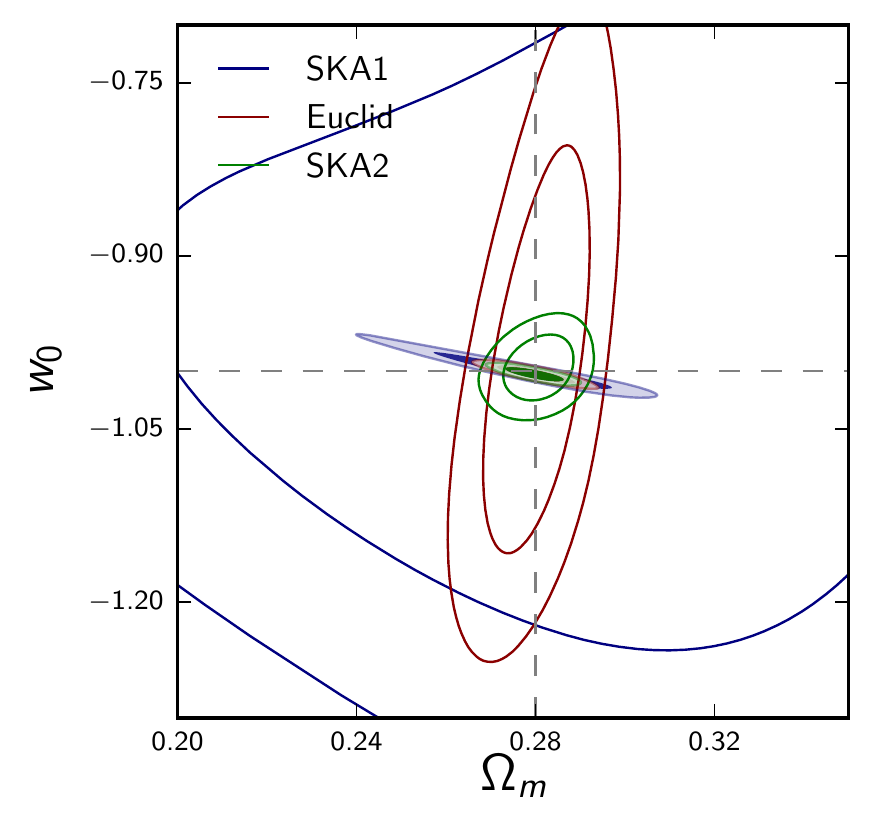}&
\includegraphics[bb=0bp 0bp 239bp 242bp,width=6cm]{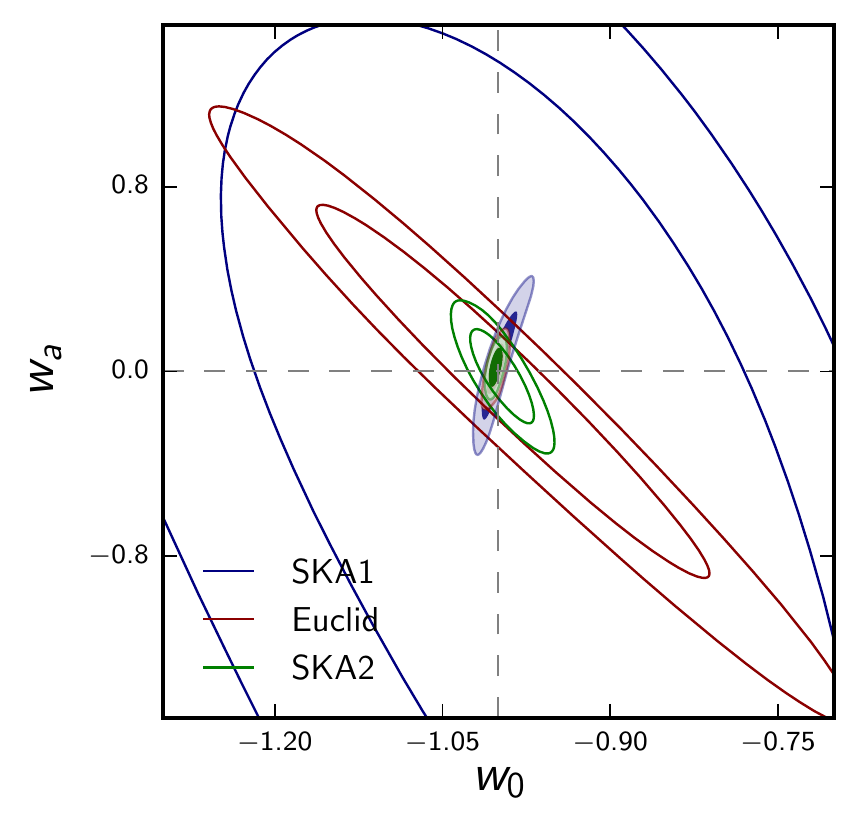}\\
 \includegraphics[bb=0bp 0bp 236bp 242bp,width=6cm]{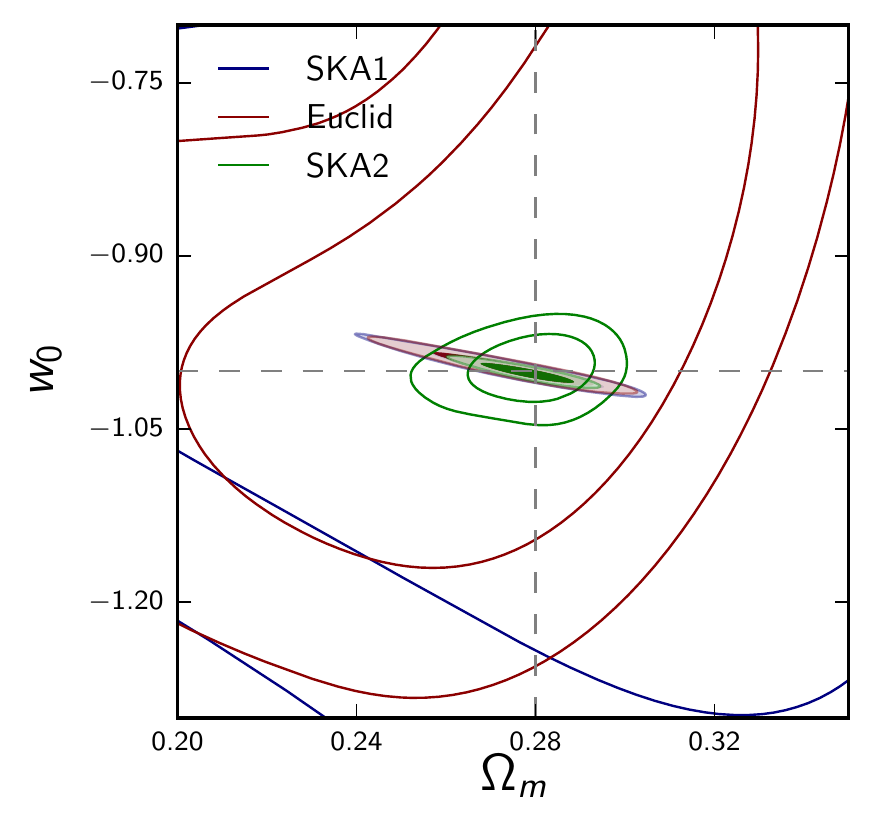}&
\includegraphics[bb=0bp 0bp 239bp 242bp,width=6cm]{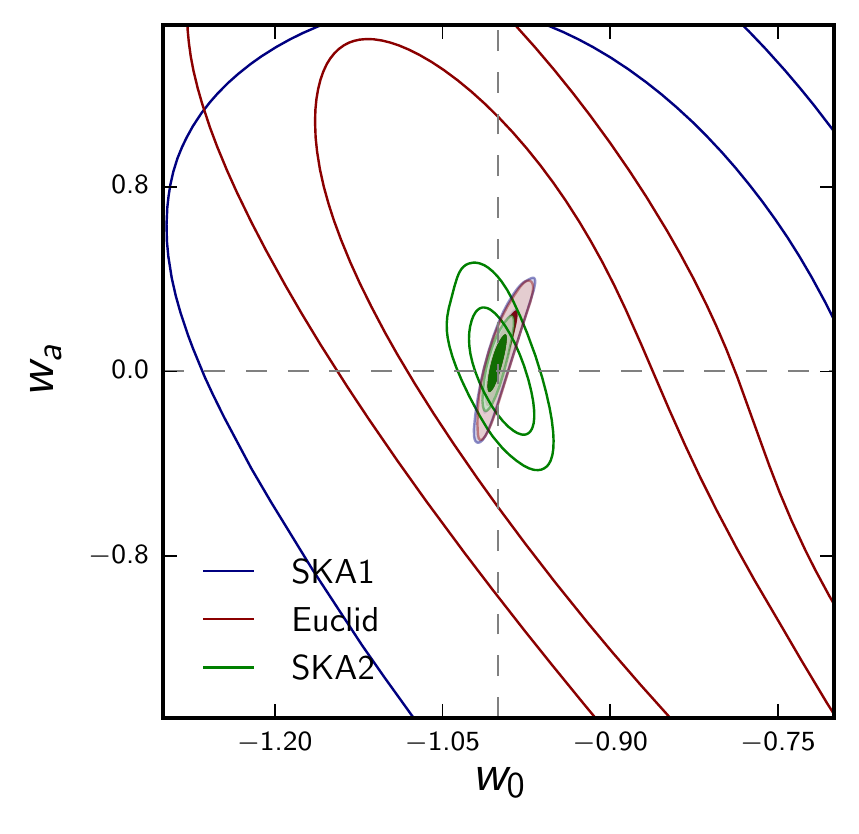}\\
\end{tabular}
\protect\protect\protect\caption{1$\sigma$ and 2$\sigma$ confidence-level contours for the standard
cosmological parameters pairs $\{\Omega_{m,0},w_{0}\}$ (left column)
and $\{w_{0},w_{a}\}$ (right column). Top panels refer to the $\Lambda$CDM
analysis, while the bottom ones describe results obtained in the MG
case. We show all the considered surveys: SKA1 (blue) Euclid (red) and SKA2 (green). 
Empty contours refer to the GC+WL combination, while full contours also include SN information.
Grey dashed lines show the fiducial values of the parameters.}

\label{fig:LCDMparams} 
\end{center}
\end{figure}

\begin{table}
\begin{center}
\begin{tabular}{|c|c|c|c|}
\hline 
% \multirow{2}{*}{Parameters }  & \multicolumn{3}{r}{FORECAST DATA} &\tabularnewline% \cline{2-5} Parameters  & SKA I  & SKA II  & EUCLID-GC  & EUCLID-FULL \tabularnewline$\Omega_{m,0}$  & $0.30\pm0.11$  & $0.279_{-0.0098}^{+0.012}$  & $0.281_{-0.0099}^{+0.013}$  & $0.2843\pm0.0065$ \tabularnewline
\textbf{Forecast data}: $\Lambda$CDM case & SKA1  & Euclid  & SKA2  \tabularnewline
\hline
$\Omega_{m,0}$ & $0.277^{+0.015}_{-0.010}$  & $0.2800\pm 0.0056$  & $0.2799^{+0.0044}_{-0.0040}$  \tabularnewline
\hline 
$w_{0}$  & $-0.9998^{+0.0074}_{-0.012}$  & $-1.0022\pm 0.0050$  & $-1.0026\pm 0.0040$  \tabularnewline
\hline 
$w_{a}$  & $0.03\pm 0.16$  & $0.011\pm 0.071$  & $0.017\pm 0.053$   \tabularnewline
\hline 
\end{tabular}

\protect

\protect\protect\protect\protect\caption{Marginalized values and $1-\sigma$ errors on the free parameters
for forecast data (which includes Euclid, SKA1 and SKA2) in the $\Lambda$CDM case.}

\label{tab:resultslambda} 
\end{center}
\end{table}

Table \ref{tab:resultslambda} and Table \ref{tab:resultsfull} report,
respectively, the results obtained in the $\Lambda$CDM limit ($Y=1$
and $\alpha=1$) and those for the full MG analysis, where both $Y$
and $\alpha$ are free parameters. These results are also shown graphically
in Fig. \ref{fig:LCDMparams} where the top panels contain the 2D
contour plots obtained in $\Lambda$CDM, while the bottom ones refer
to the MG case.

\begin{figure}[!h]
\begin{center}
\begin{tabular}{cc}
\includegraphics[bb=0bp 0bp 236bp 242bp,width=6cm]{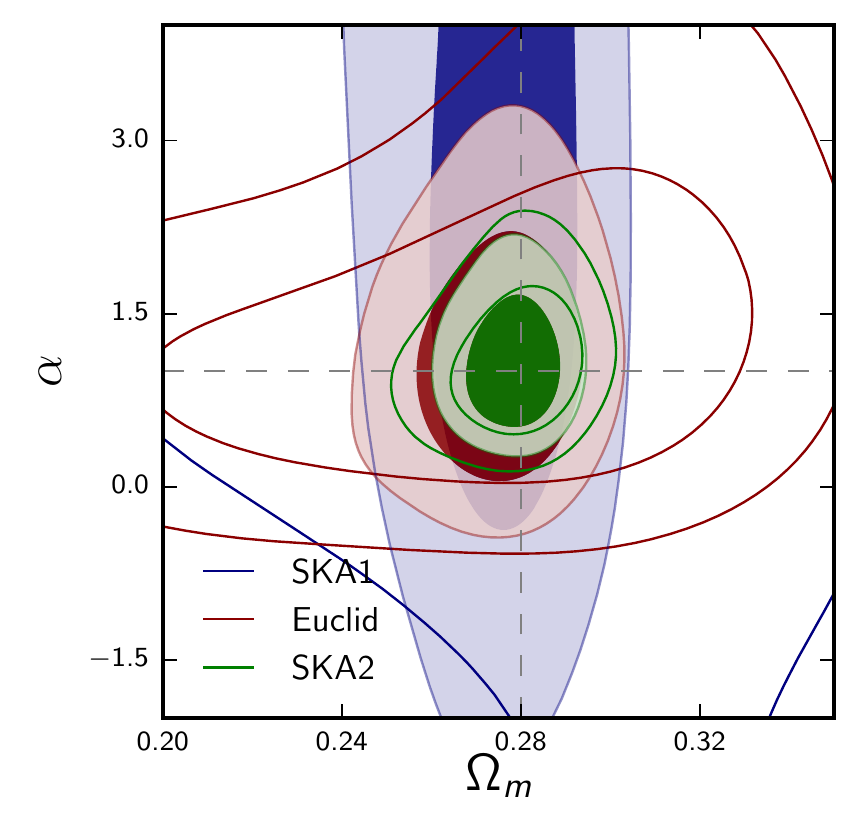}&
\includegraphics[bb=0bp 0bp 239bp 242bp,width=6cm]{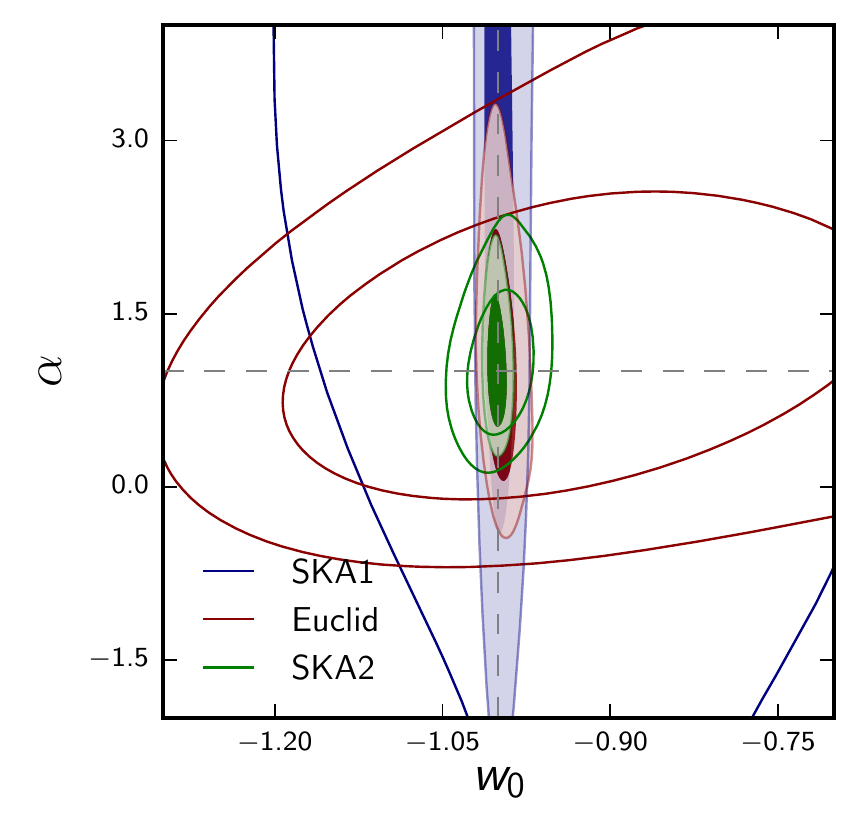}\\
\includegraphics[bb=0bp 0bp 236bp 242bp,width=6cm]{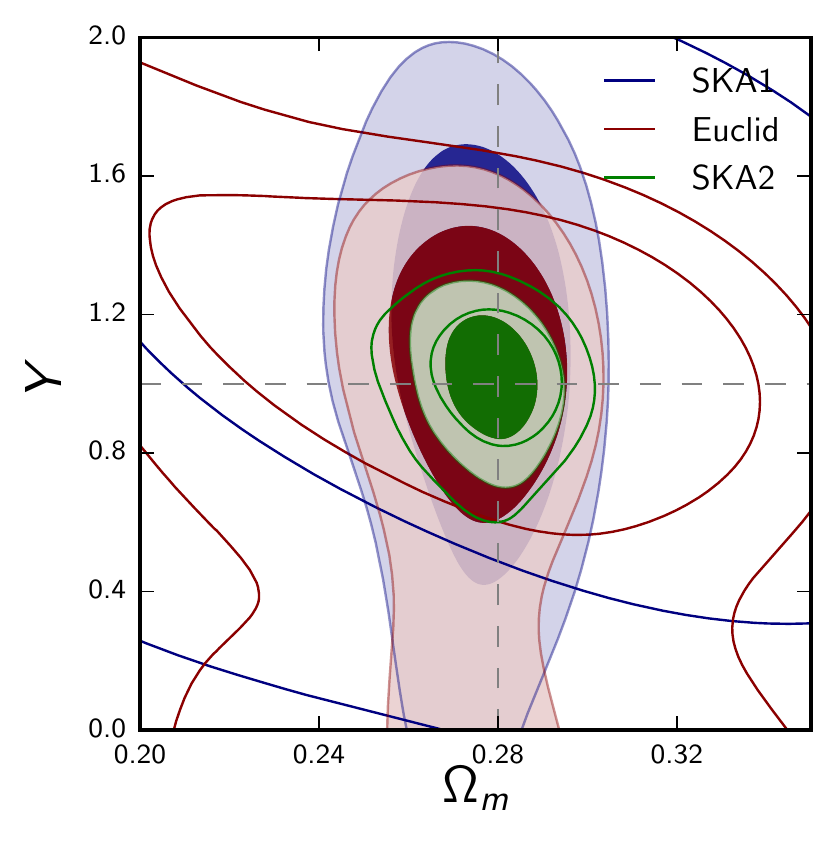}&
\includegraphics[bb=0bp 0bp 239bp 242bp,width=6cm]{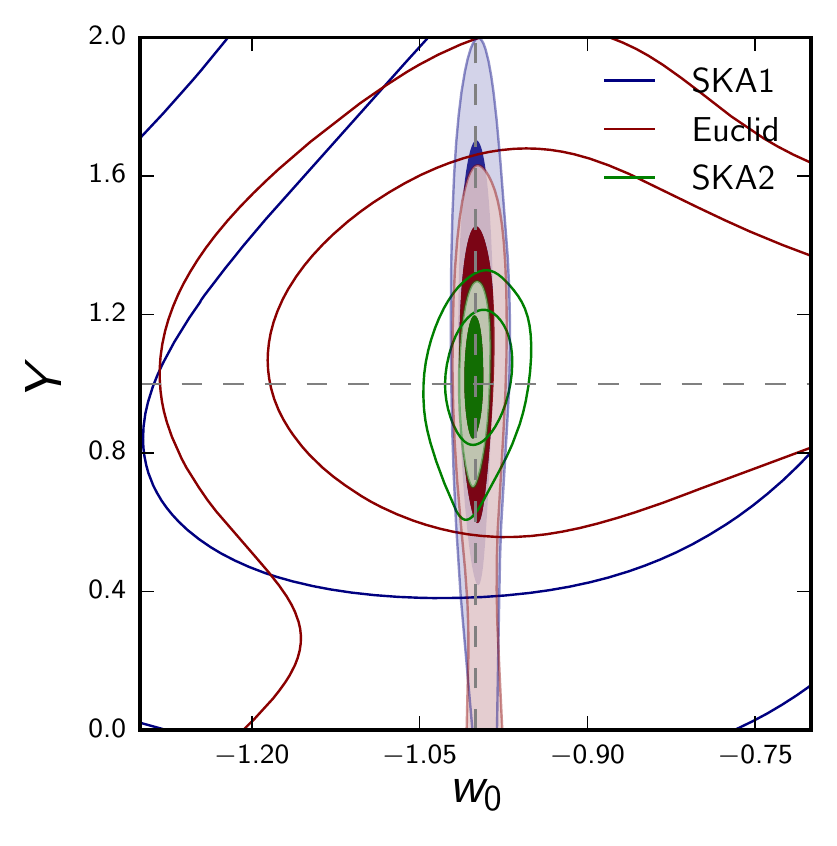}\\
\end{tabular}
 \protect\protect\protect\caption{1$\sigma$ and 2$\sigma$ confidence-level contours for the parameters
$\{\Omega_{m,0},\alpha\}$ (top left), $\{w_{0},\alpha\}$ (top right),
$\{\Omega_{m,0},Y\}$ (bottom left) and $\{w_{0},Y\}$ (bottom right).
We show all the considered surveys: SKA1 (blue) Euclid (red) and SKA2 (green).
Empty contours refer to the GC+WL combination, while full contours also include SN information. 
Grey dashed lines show the fiducial values of the parameters.}

\label{fig:noLCDM} 
\end{center}
\end{figure}

It is immediate to see how, although all the considered data combinations
are expected to be extremely precise in measuring the expansion history
of the Universe, constraints brought by galaxy clustering and weak lensing
on the standard cosmological parameters are degraded when MG parameters
are included in the analysis; this can be noticed comparing the empty contours
of the top and bottom panels of Fig. \ref{fig:LCDMparams}. 
In the Euclid case for example, the 
errors on $\Omega_{m,0}$, $w_{0}$ and $w_{a}$ increase respectively by roughly
$5$, $2$ and $3$ times.
This is due, as already discussed for
currently available data in the previous section, to the degeneracies
between standard parameters and $Y$ and $\alpha$, which are shown
by the empty contours of Fig. \ref{fig:noLCDM}. The net effect of allowing modified gravity
is therefore to make more expansion histories viable despite the extreme
sensitivity of future surveys.

However, as SN information is not affected by the MG parameters, which modify only 
the linear perturbations, its inclusion strongly helps to break degeneracies, as 
it can be seen in the filled contours of Fig. \ref{fig:noLCDM} and in Table \ref{tab:resultsfull}.\\
It is interesting to notice how the current constraints on MG parameters
significantly improve using forecasts for these experiments; 
this means that despite the degeneracies highlighted
here, future surveys will be able to significantly reduce the parameter
space regions allowed for modified gravity theories.

\begin{table}
\begin{center}
\begin{tabular}{|c|c|c|c|c|}

\hline 
% \multirow{2}{*}{Parameters }  & \multicolumn{3}{r}{FORECAST DATA} &\tabularnewline% \cline{2-5} Parameters  & SKA I  & SKA II  & EUCLID-GC  & EUCLID-FULL \tabularnewline$Y$  & $1.12_{-0.98}^{+0.34}$  & $1.22_{-0.45}^{+0.38}$  & $1.18\pm0.47$  & $1.07_{-0.21}^{+0.29}$ \tabularnewline
\textbf{Forecast data: full case}& SKA1  & Euclid  & SKA2\tabularnewline
\hline 
$\Omega_{m,0}$  & $0.275^{+0.014}_{-0.011}$  & $0.276^{+0.012}_{-0.0095}$  & $0.2782^{+0.0072}_{-0.0062}$ \tabularnewline
\hline 
$\alpha$  & $> -0.0431$  & $1.17^{+0.61}_{-1.0}$  & $1.12^{+0.30}_{-0.45}$ \tabularnewline
\hline 
$w_{0}$  & $-0.9986^{+0.0084}_{-0.012}$  & $-0.9987^{+0.0069}_{-0.0098}$  & $-1.0013^{+0.0051}_{-0.0061}$ \tabularnewline
\hline 
$w_{a}$  & $0.06\pm 0.15$  & $0.05\pm 0.13$  & $0.036\pm 0.084$ \tabularnewline
\hline 
$Y$      & $1.09^{+0.45}_{-0.32}$  & $0.98^{+0.35}_{-0.14}$  & $1.02^{+0.13}_{-0.11}$ \tabularnewline
\hline
\end{tabular}

\protect

\protect\protect\protect\protect\caption{Marginalized values and $1-\sigma$ errors on the free parameters
for forecast data (which includes Euclid, SKA1 and SKA2) when MG parameters are allowed to vary.}

\label{tab:resultsfull} 
\end{center}
\end{table}

\begin{figure}[!h]
\begin{center}
\includegraphics[bb=0bp 0bp 236bp 242bp,width=8cm]{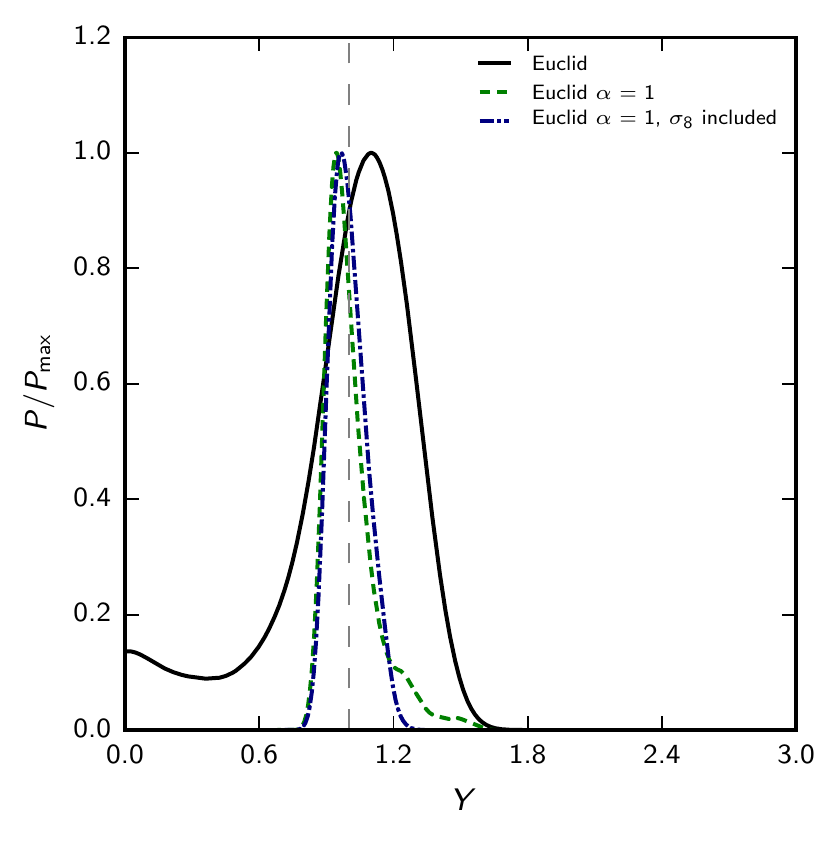}
 \protect\protect\protect\caption{Posterior distribution for the $Y$ parameter recovered marginalizing over $\sigma_8$ and varying both $Y$ and $\alpha$ (black solid line) or fixing $\alpha=1$ (green dashed line). The posterior is also shown with a free to vary $\sigma_8$ without marginalizing it out of the likelihood (blue dot-dashed line).}

\label{fig:margeff} 
\end{center}
\end{figure}

Figure \ref{fig:margeff} shows the posterior distribution on $Y$ obtained using the Euclid full dataset (i.e. with the inclusion of supernovae) both with a varying $\alpha$ and $\alpha$ fixed to $1$. In this second case we also show the posterior obtained without marginalizing a priori on $\sigma_8$; in this case we use the $\chi^2$ of Eq. (\ref{eq:chifull}), with a free-to-vary  $\sigma_8$ derived from the cosmological parameters assuming GR. In this plot we notice, as expected, how the constraint on $Y$ are much tighter when $\alpha=1$ with respect to the general initial conditions case. We notice also how the inclusion of $\sigma_8$ in the analysis slightly improves further the constraint on $Y$ although relying on a specific assumption of the gravity theory.

\section{A cosmological exclusion plot}

\label{sec:vi}

In this section we wish to obtain an exclusion plot, i.e. the region
of the parameter space that a future redshift surveys can achieve,
in analogy to what already done in \cite{Taddei:2014wqa}. We remind
that, in the quasi-static limit of the Horndeski theory, the expression
for $Y$ is given by Eq. (\ref{Y1-1}). If we perform a Fourier anti-transformation
of this equation, we obtain a Yukawa-like gravitational potential
\begin{equation}
\Psi(r)=-\frac{G_{0}M}{r}h_{1}\left(1+Qe^{-r/\lambda}\right)
\end{equation}
where $h_{5}=(1+Q)\lambda^{2}$ and $h_{3}=\lambda^{2}$ (notice that
$Mh_{1}$ is the observable, not $h_{1}$ alone), and where the mass
$M$ is to be interpreted as the gravitational mass of a source at
the origin. Here $G_{0}$ is the gravitational constant one would
measure in laboratory where the effects of the modification of gravity
are assumed to be screened or baryons to be uncoupled. Now, we want to solve Eq. (\ref{eq:a-1-1})
in the quasi-static Horndeski result coupled to the background defined
in Eq. (\ref{eq:back}), so in total, we have seven parameters: the
strength $Q$ and the range $\lambda$ of the Yukawa term (we assume
that they are constant in the observed range), $h_{1}$, the initial
conditions $\alpha$ and $\Omega_{m,0},w_{0},w_{a.}$. If one were to satisfy the present-time constraints
on $\dot{(GM)}/GM$ at all times, then $h_1=1$ with great precision. As  we already remarked, however, this extrapolation of present constraints 
to the past is unwarranted and we leave therefore $h_1$ as a free parameter to be marginalized over. 

We reproduce the exclusion plot only for the Euclid and SKA2 cases since, as we have seen before, they are those that
best constrain the parameters. In order to solve Eq. (\ref{eq:a-1-1}),
we need to have a minimum of three $k$-bins for every value of the
redshift. We take the minimum binning value of $k$ as $k_{min}=0.007$
$h/$Mpc (the results are very much insensitive to this choice) and
the value of the highest $k$ is conservatively chosen to be below
the scale of non-linearity at the redshift of the first bin for the
Euclid case. For simplicity, we assume that the $k$-values are the
same for all the redshift bins ($k_{min}-k_{1}:  0.007-0.022$, $k_{1}-k_{2}: 0.022-0.063$, $k_{2}-k_{max}=0.063-0.180$) for both surveys. 
In Table \ref{tab7} we display the fiducial values
and errors for $f\sigma_{8}$ and $E$ at every redshift and every
$k$-bin for the Euclid case and in Table \ref{tab8}
for the SKA2-full. We numerically solve Eq. (\ref{eq:a-1-1}) inserting
the value of $k$ corresponding to the central value of each $k$-bin and
we vary the value of $\lambda$ from 0 to 120 Mpc$/h$; for every
step, we calculate the errors on $Q$ at 68\% and 95\% of c.l., marginalizing
over all the other parameters. The region which is outside the errors
on $Q$ is therefore the region that a future survey will be able
to rule out. The parameters $Q$ and $\lambda$ are the cosmological
analog of the parameters employed in laboratory experiments to test
deviations from Newtonian gravity, see e.g. \cite{2007PhRvL..98b1101K}.

\begin{figure}[!htb]
\begin{centering}
\hspace*{-1cm} %
\begin{tabular}{cc}
\includegraphics[width=8cm]{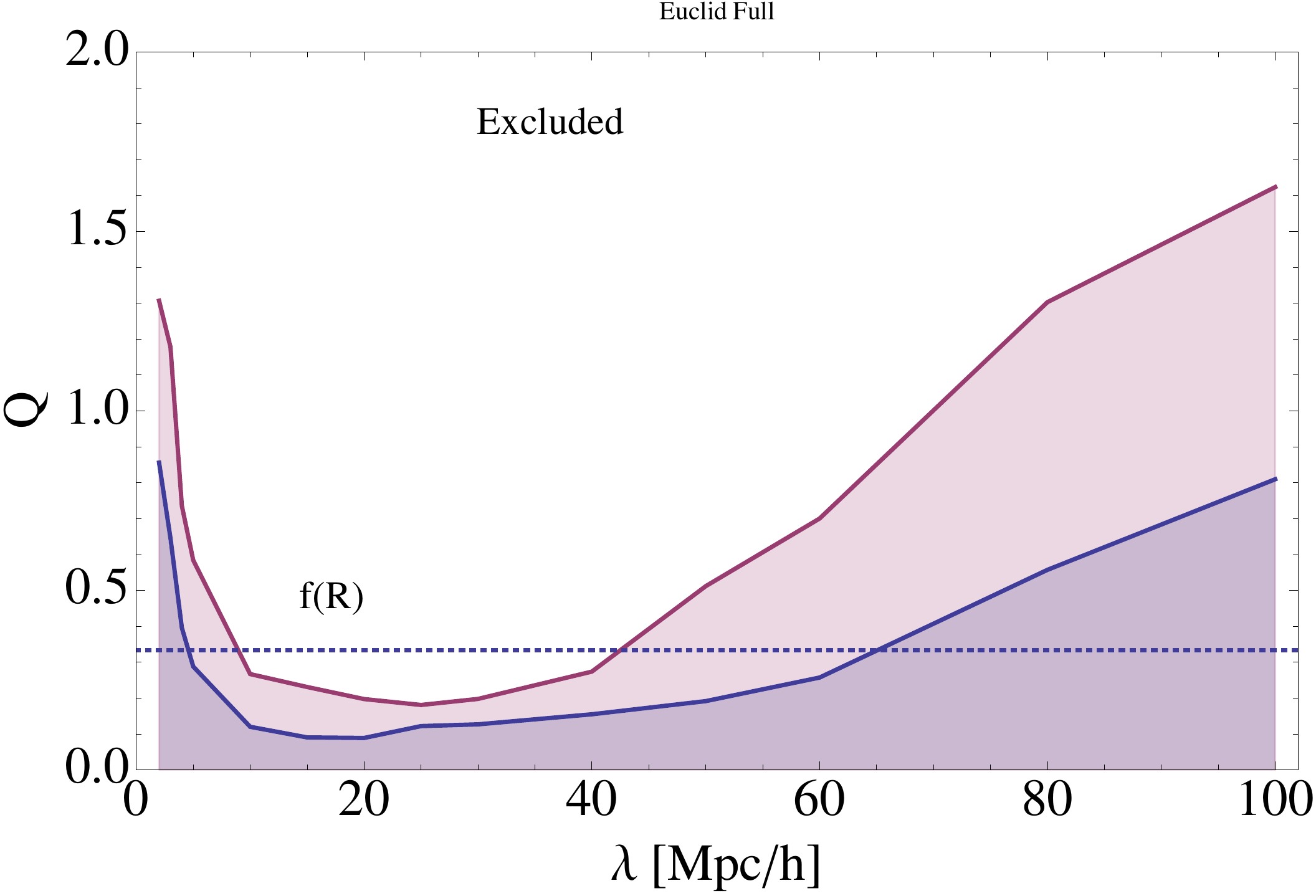}\includegraphics[width=8cm]{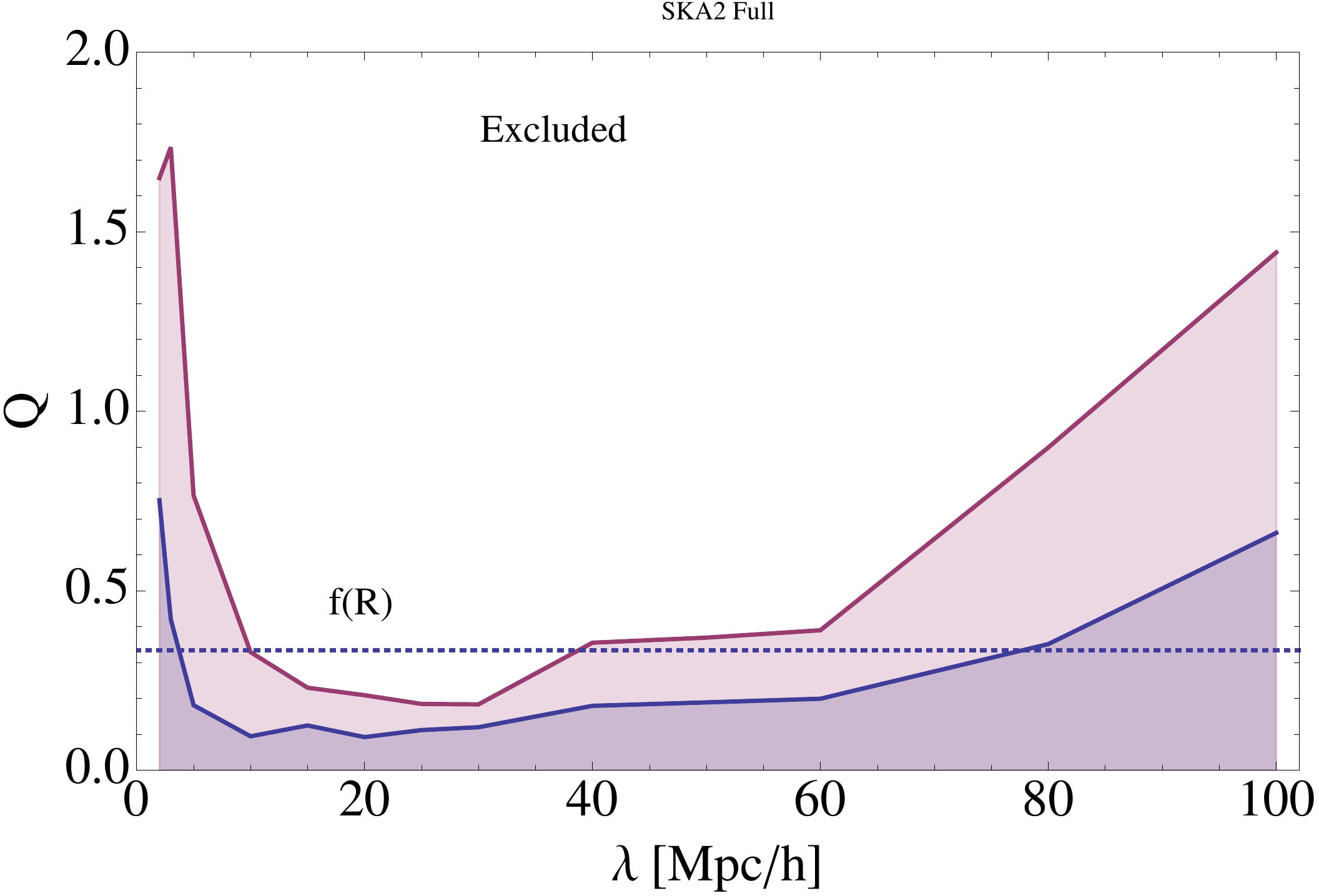}  & \tabularnewline
\end{tabular}\protect\protect\caption{Forecast of a cosmological exclusion plot for the Euclid (left panel) and 
for the SKA2 (right panel) surveys. In both cases, the
darker region is the 68\% c.l. region, the lighter one is the 95\% c.l. region. The dotted line 
is the value of $Q$ in $f(R)$ models.}

\par
\label{fig:explot} 
\end{centering}
\end{figure}

Fig. \ref{fig:explot} shows the region that the Euclid survey (left panel)
and  SKA2 survey (right panel) can exclude. It appears that they are very similar. For very small values of $\lambda$, the 
strength $Q$ is unconstrained, since in this limit the fifth-force effects decay at small distances.
$Q$ is weakly constrained also at very large $\lambda$ since in this limit $Y$ becomes scale independent and the overall constant $h_1$ is marginalised over. 
The constraints are therefore maximized at intermediate values $10 Mpc/h < \lambda< 50 Mpc/h$. In this regime, both  surveys will constrain $Q$ to be smaller than 0.2-0.3 (68\% c.l.)

In the popular class of $f(R)$ models, one has $Q=1/3$ and the range $\lambda=1/m$ where $m$ is the mass of the scalaron given by
\begin{equation}
m^2=\frac{R}{3}(\frac{1+f_{,R}}{Rf_{,RR}}-1)
\end{equation}
where $f_{,R},f_{,RR}$ are derivatives with respect to $R$.
Moreover,
$h_1=1/(1+f_{,R})$. Often constraints on $f(R)$ are obtained for specific models and fixing the background to be $\Lambda$CDM,
 \cite{Dossett:2014oia,Hojjati:2015ojt}. 
Our exclusion plot allows us to remove  these assumptions since we do not fix the modified gravity model and 
we marginalize over the background.
From Fig. \ref{fig:explot} we see  that $f(R)$ models can be ruled out only if the range $\lambda=1/m$
is within $10-40$ Mpc/$h$ at $95\%$ c.l. and $5-70$ Mpc/$h$ at $68\%$ c.l. .

\begin{table}
\begin{center}
\begin{tabular}{|c|c|c|c|c|c|}
\hline 
$\bar{{z}}$  & $i$  & $f\sigma_{8}(z)$  & $\Delta f\sigma_{8}(z)$  & $E$  & $\Delta E$\tabularnewline
\hline 
\hline 
\multirow{3}{*}{0.6}  & 1  & \multirow{3}{*}{0.469 }  & 0.069    & \multirow{3}{*}{1.37}  & \multirow{3}{*}{0.003}\tabularnewline
\hline 
                      & 2  &                          & 0.017   &  & \tabularnewline
\hline 
                      & 3  &                          & 0.009  &  & \tabularnewline
\hline 
\multirow{3}{*}{0.8}  & 1  & \multirow{3}{*}{0.457}  & 0.005  & \multirow{3}{*}{1.53}  & \multirow{3}{*}{0.03}\tabularnewline
\hline 
 & 2  &  & 0.013  &  & \tabularnewline
\hline 
 & 3  &  & 0.007  &  & \tabularnewline
\hline 
\multirow{3}{*}{1.0}  & 1  & \multirow{3}{*}{0.438 }  & 0.044   & \multirow{3}{*}{1.72}  & \multirow{3}{*}{0.06}\tabularnewline
\hline 
 & 2  &  & 0.011    &  & \tabularnewline
\hline 
 & 3  &  & 0.005    &  & \tabularnewline
\hline 
\multirow{3}{*}{1.2}  & 1  & \multirow{3}{*}{0.417}  & 0.039   & \multirow{3}{*}{1.92}  & \multirow{3}{*}{0.010}\tabularnewline
\hline 
 & 2  &  & 0.009    &  & \tabularnewline
\hline 
 & 3  &  & 0.004   &  & \tabularnewline
\hline 
\multirow{3}{*}{1.4}  & 1  & \multirow{3}{*}{0.396}  & 0.035   & \multirow{3}{*}{2.14}  & \multirow{3}{*}{0.018}\tabularnewline
\hline 
 & 2  &  & 0.008  &  & \tabularnewline
\hline 
 & 3  &  & 0.004    &  & \tabularnewline
\hline 
\multirow{3}{*}{1.8 }  & 1  & \multirow{3}{*}{0.354 }  & 0.019  & \multirow{3}{*}{2.62}  & \multirow{3}{*}{0.035}\tabularnewline
\hline 
 & 2  &  & 0.005   &  & \tabularnewline
\hline 
 & 3  &  & 0.003    &  & \tabularnewline
\hline 
\end{tabular}\protect\protect\protect\protect\protect\protect\caption{Fiducial values and relative errors for $f\sigma_{8}$ data at every
redshift $\bar{{z}}$ and every $k$-bin (labeled with the index $i$,
from \cite{Amendola:2013qna}), for the Euclid survey.}
\label{tab7} 
\end{center}
\end{table}

\begin{table}
\begin{center}
\begin{tabular}{|c|c|c|c|c|c|}
\hline 
$\bar{{z}}$  & $i$  & $f\sigma_{8}(z)$  & $\Delta$ $f\sigma_{8}(z)$   & $E$  & $\Delta E$\tabularnewline
\hline 
\hline 
\multirow{3}{*}{0.26}  & 1  & \multirow{3}{*}{0.459}  & 0.060  & \multirow{3}{*}{1.13}  & \multirow{3}{*}{0.0003}\tabularnewline
\hline 
 & 2  &  & 0.015  &  & \tabularnewline
\hline 
 & 3  &  & 0.010 &  & \tabularnewline
\hline 
\multirow{3}{*}{0.75}  & 1  & \multirow{3}{*}{0.460}  & 0.030   & \multirow{3}{*}{1.48}  & \multirow{3}{*}{0.0009}\tabularnewline
\hline 
 & 2  &  & 0.007  &  & \tabularnewline
\hline 
 & 3  &  & 0.004   &  & \tabularnewline
\hline 
\multirow{3}{*}{1.25}  & 1  & \multirow{3}{*}{0.412}  & 0.025   & \multirow{3}{*}{1.97}  & \multirow{3}{*}{0.0043}\tabularnewline
\hline 
 & 2  &  & 0.006    &  & \tabularnewline
\hline 
 & 3  &  & 0.003    &  & \tabularnewline
\hline 
\multirow{3}{*}{1.75}  & 1  & \multirow{3}{*}{0.359}  & 0.026    & \multirow{3}{*}{2.55}  & \multirow{3}{*}{0.016}\tabularnewline
\hline 
 & 2  &  & 0.006   &  & \tabularnewline
\hline 
 & 3  &  & 0.003   &  & \tabularnewline
\hline 
\multirow{3}{*}{2.25}  & 1  & \multirow{3}{*}{0.314}  & 0.038    & \multirow{3}{*}{3.21}  & \multirow{3}{*}{0.020}\tabularnewline
\hline 
 & 2  &  & 0.011   &  & \tabularnewline
\hline 
 & 3  &  & 0.008  &  & \tabularnewline
\hline 
\multirow{3}{*}{2.75}  & 1  & \multirow{3}{*}{0.277}  & 0.12   & \multirow{3}{*}{3.93}  & \multirow{3}{*}{0.095}\tabularnewline
\hline 
 & 2  &  & 0.049   &  & \tabularnewline
\hline 
 & 3  &  & 0.044   &  & \tabularnewline
\hline 
\end{tabular}\protect\protect\protect\protect\protect\protect\caption{Fiducial values and relative errors for $f\sigma_{8}$ data at every
redshift $\bar{{z}}$ and every $k$-bin (labeled with the index $i$),
for the SKA2 survey.}

\label{tab8} 
\end{center}
\end{table}

\section{Conclusions}\label{sec:vii} 
In this paper we investigated deviations from the standard Poisson's equation in terms of general modifications of the gravitational theory, parametrized through 
an effective gravitational constant $Y$. With respect to previous results we make a further attempt 
towards a model independent approach, using general initial conditions, encoded in the $\alpha$ parameter, and
marginalizing $\sigma_8$ out of the likelihood, thus removing the need of a choice of the shape of the power
spectrum. We assumed the parameters $Y$ and $\alpha$ to be constant or, 
in the case of $Y$, to follow a Horndeski behavior. Alongside the modified growth of perturbations, we also assume a general expansion history described by
$w(a)=w_0+w_a(1-a)$.

As explained in the Introduction and in the forecast section, one should be careful in comparing our results to similar analyses: generally speaking, the more one tries to be model-independent, the more is likely to obtain weaker constraints. These constraints are however by construction more robust in the sense that they stay the same even when other models of, e.g., initial conditions or bias, are assumed.

In the constant parameters assumption, we analyzed this general parametrization with currently available supernovae and galaxy surveys, investigating the effect of deviations 
from the standard growth and  expansion history. In order to preserve the generality of our approach, we marginalize an overall amplitude out of the data when computing our likelihood, 
as this depends on unknown initial conditions.
We find that current data are not able to constrain the modified gravity parameters with good precision, as $\alpha$ is constrained with a $\approx300\%$ error, while only
an upper bound can be obtained on $Y$. Interestingly, we find that the growth-rate data prefer $\Omega_{m,0}$ to assume small values, while supernovae data support a maximum 
likelihood $\Omega_{m,0}\approx0.3$. The combined analysis still prefers low values of the matter density, but the effect of supernovae is to make more likely  values 
excluded by the growth-data-only analysis, in the modified gravity case.

We also forecast the performances of two future experiments: SKA, in the SKA1 and SKA2 configurations, and Euclid. 
In both cases we combine the expected information coming from galaxy clustering and weak lensing, and we include also a forecast for a future SN survey.
It is possible to notice how these future experiments will significantly improve
our knowledge of gravity, as possible deviations from the standard GR will be constrained with a $\approx30\%$ error on $Y$ and a $\approx70\%$ 
 one on $\alpha$ in the most sensitive case considered here (SKA2).

Finally we adopt a Horndeski form for $Y$, i.e. a Yukawa-like correction to the Newtonian
gravitational potential. The constraints imposed by future experiments are represented as
a cosmological exclusion plot between the parameters $Q$ and $\lambda$, i.e. the strength and range of the Yukawa force introduced by the modification of gravity. 
We find that the strength will be constrained to be less than  0.2 at 68\% c.l. 
of the Newtonian gravitational strength for both the Euclid and the SKA2 cases if the interaction 
range is around 10-50 Mpc/$h$.
This constraint will be sufficient to rule out all $f(R)$ models with a range in this regime.

\section{Acknowledgments}
L. Taddei thanks ``Fondazione Angelo della Riccia" for financial support. We thank DFG for support through the Transregio 33 project "The Dark Universe".

 \bibliographystyle{unsrt}
\bibliography{observables,amendola,massive-gravity,growth-rate}

\begin{thebibliography}{10}

\bibitem{Riess:1998cb}
Adam~G. Riess et~al.
\newblock {Observational evidence from supernovae for an accelerating universe
  and a cosmological constant}.
\newblock {\em Astron. J.}, 116:1009--1038, 1998.

\bibitem{Perlmutter:1998np}
S.~Perlmutter et~al.
\newblock {Measurements of Omega and Lambda from 42 high redshift supernovae}.
\newblock {\em Astrophys. J.}, 517:565--586, 1999.

\bibitem{Amendola2010}
Luca Amendola and Shinji Tsujikawa.
\newblock {\em {Dark Energy: Theory and Observations}}.
\newblock Cambridge University Press, 2010.

\bibitem{Euclid_TWG}
L.~{Amendola}, S.~{Appleby}, D.~{Bacon}, T.~{Baker}, M.~{Baldi}, N.~{Bartolo},
  A.~{Blanchard}, C.~{Bonvin}, S.~{Borgani}, E.~{Branchini}, C.~{Burrage},
  S.~{Camera}, C.~{Carbone}, L.~{Casarini}, M.~{Cropper}, C.~{deRham}, C.~{di
  Porto}, A.~{Ealet}, P.~G. {Ferreira}, F.~{Finelli}, J.~{Garcia-Bellido},
  T.~{Giannantonio}, L.~{Guzzo}, A.~{Heavens}, L.~{Heisenberg}, C.~{Heymans},
  H.~{Hoekstra}, L.~{Hollenstein}, R.~{Holmes}, O.~{Horst}, K.~{Jahnke}, T.~D.
  {Kitching}, T.~{Koivisto}, M.~{Kunz}, G.~{La Vacca}, M.~{March},
  E.~{Majerotto}, K.~{Markovic}, D.~{Marsh}, F.~{Marulli}, R.~{Massey},
  Y.~{Mellier}, D.~F. {Mota}, N.~{Nunes}, W.~{Percival}, V.~{Pettorino},
  C.~{Porciani}, C.~{Quercellini}, J.~{Read}, M.~{Rinaldi}, D.~{Sapone},
  R.~{Scaramella}, C.~{Skordis}, F.~{Simpson}, A.~{Taylor}, S.~{ Thomas},
  R.~{Trotta}, L.~{Verde}, F.~{Vernizzi}, A.~{Vollmer}, Y.~{Wang}, J.~{Weller},
  and T.~{Zlosnik}.
\newblock {Cosmology and fundamental physics with the Euclid satellite}.
\newblock {\em ArXiv e-prints}, June 2012.

\bibitem{2013PhRvD..87b3501A}
L.~{Amendola}, M.~{Kunz}, M.~{Motta}, I.~D. {Saltas}, and I.~{Sawicki}.
\newblock {Observables and unobservables in dark energy cosmologies}.
\newblock {\em \prd}, 87(2):023501, January 2013.

\bibitem{Amendola:2013qna}
Luca Amendola, Simone Fogli, Alejandro Guarnizo, Martin Kunz, and Adrian
  Vollmer.
\newblock {Model-independent constraints on the cosmological anisotropic
  stress}.
\newblock {\em Phys.Rev.}, D89:063538, 2014.

\bibitem{DeFelice:2011hq}
Antonio De~Felice, Tsutomu Kobayashi, and Shinji Tsujikawa.
\newblock {Effective gravitational couplings for cosmological perturbations in
  the most general scalar-tensor theories with second-order field equations}.
\newblock {\em Phys.Lett.}, B706:123--133, 2011.

\bibitem{Amendola:2012ky}
Luca Amendola, Martin Kunz, Mariele Motta, Ippocratis~D. Saltas, and Ignacy
  Sawicki.
\newblock {Observables and unobservables in dark energy cosmologies}.
\newblock {\em Phys.Rev.}, D87:023501, 2013.

\bibitem{Silvestri:2013ne}
Alessandra Silvestri, Levon Pogosian, and Roman~V. Buniy.
\newblock {A practical approach to cosmological perturbations in modified
  gravity}.
\newblock 2013.

\bibitem{Sawicki:2015zya}
Ignacy Sawicki and Emilio Bellini.
\newblock {Limits of quasistatic approximation in modified-gravity
  cosmologies}.
\newblock {\em Phys. Rev.}, D92(8):084061, 2015.

\bibitem{Taddei:2014wqa}
Laura Taddei and Luca Amendola.
\newblock {A cosmological exclusion plot: Towards model-independent constraints
  on modified gravity from current and future growth rate data}.
\newblock {\em JCAP}, 1502(02):001, 2015.

\bibitem{Chevallier_Polarski_2001}
Michel Chevallier and David Polarski.
\newblock {Accelerating universes with scaling dark matter}.
\newblock {\em Int.J.Mod.Phys.}, D10:213--224, 2001.

\bibitem{Lewis:2002ah}
Antony Lewis and Sarah Bridle.
\newblock {Cosmological parameters from CMB and other data: A Monte Carlo
  approach}.
\newblock {\em Phys. Rev.}, D66:103511, 2002.

\bibitem{Betoule:2014frx}
M.~Betoule et~al.
\newblock {Improved cosmological constraints from a joint analysis of the
  SDSS-II and SNLS supernova samples}.
\newblock {\em Astron.Astrophys.}, 568:A22, 2014.

\bibitem{Macaulay:2013swa}
Edward Macaulay, Ingunn~Kathrine Wehus, and Hans~Kristian Eriksen.
\newblock {A Lower Growth Rate from Recent Redshift Space Distortions than
  Expected from Planck}.
\newblock 2013.

\bibitem{More:2014uva}
Surhud More, Hironao Miyatake, Rachel Mandelbaum, Masahiro Takada, David
  Spergel, et~al.
\newblock {The Weak Lensing Signal and the Clustering of BOSS Galaxies:
  Cosmological Constraints}.
\newblock 2014.

\bibitem{Beutler:2012px}
Florian Beutler, Chris Blake, Matthew Colless, D.~Heath Jones, Lister
  Staveley-Smith, et~al.
\newblock {The 6dF Galaxy Survey: z ~ 0 measurement of the growth rate and
  sigma-8}.
\newblock {\em Mon.Not.Roy.Astron.Soc.}, 423:3430--3444, 2012.

\bibitem{Samushia:2011cs}
Lado Samushia, Will~J. Percival, and Alvise Raccanelli.
\newblock {Interpreting large-scale redshift-space distortion measurements}.
\newblock {\em Mon.Not.Roy.Astron.Soc.}, 420:2102--2119, 2012.

\bibitem{Tojeiro:2012rp}
Rita Tojeiro, W.J. Percival, J.~Brinkmann, J.R. Brownstein, D.~Eisenstein,
  et~al.
\newblock {The clustering of galaxies in the SDSS-III Baryon Oscillation
  Spectroscopic Survey: measuring structure growth using passive galaxies}.
\newblock {\em Mon.Not.Roy.Astron.Soc.}, 424:2339--2344, 2012.

\bibitem{Blake:2012pj}
Chris Blake, Sarah Brough, Matthew Colless, Carlos Contreras, Warrick Couch,
  et~al.
\newblock {The WiggleZ Dark Energy Survey: Joint measurements of the expansion
  and growth history at z < 1}.
\newblock {\em Mon.Not.Roy.Astron.Soc.}, 425:405--414, 2012.

\bibitem{delaTorre:2013rpa}
S.~de~la Torre, L.~Guzzo, J.A. Peacock, E.~Branchini, A.~Iovino, et~al.
\newblock {The VIMOS Public Extragalactic Redshift Survey (VIPERS). Galaxy
  clustering and redshift-space distortions at z=0.8 in the first data
  release}.
\newblock 2013.

\bibitem{Percival:2004fs}
Will~J. Percival et~al.
\newblock {The 2dF Galaxy Redshift Survey: Spherical harmonics analysis of
  fluctuations in the final catalogue}.
\newblock {\em Mon.Not.Roy.Astron.Soc.}, 353:1201, 2004.

\bibitem{Song:2008qt}
Yong-Seon Song and Will~J. Percival.
\newblock {Reconstructing the history of structure formation using Redshift
  Distortions}.
\newblock {\em JCAP}, 0910:004, 2009.

\bibitem{Chuang:2012qt}
Chia-Hsun Chuang and Yun Wang.
\newblock {Modeling the Anisotropic Two-Point Galaxy Correlation Function on
  Small Scales and Improved Measurements of $H(z)$, $D_A(z)$, and $\beta(z)$
  from the Sloan Digital Sky Survey DR7 Luminous Red Galaxies}.
\newblock {\em Mon.Not.Roy.Astron.Soc.}, 435:255--262, 2013.

\bibitem{Chuang:2013wga}
Chia-Hsun Chuang, Francisco Prada, Florian Beutler, Daniel~J. Eisenstein,
  Stephanie Escoffier, et~al.
\newblock {The clustering of galaxies in the SDSS-III Baryon Oscillation
  Spectroscopic Survey: single-probe measurements from CMASS and LOWZ
  anisotropic galaxy clustering}.
\newblock 2013.

\bibitem{Beutler:2013yhm}
Florian Beutler et~al.
\newblock {The clustering of galaxies in the SDSS-III Baryon Oscillation
  Spectroscopic Survey: Testing gravity with redshift-space distortions using
  the power spectrum multipoles}.
\newblock 2013.

\bibitem{Samushia:2013yga}
Lado Samushia, Beth~A. Reid, Martin White, Will~J. Percival, Antonio~J. Cuesta,
  et~al.
\newblock {The Clustering of Galaxies in the SDSS-III Baryon Oscillation
  Spectroscopic Survey (BOSS): measuring growth rate and geometry with
  anisotropic clustering}.
\newblock 2013.

\bibitem{Reid:2014iaa}
Beth~A. Reid, Hee-Jong Seo, Alexie Leauthaud, Jeremy~L. Tinker, and Martin
  White.
\newblock {A 2.5% measurement of the growth rate from small-scale redshift
  space clustering of SDSS-III CMASS galaxies}.
\newblock 2014.

\bibitem{Ade:2015xua}
P.~A.~R. Ade et~al.
\newblock {Planck 2015 results. XIII. Cosmological parameters}.
\newblock 2015.

\bibitem{Euclid-r}
R.~{Laureijs}, J.~{Amiaux}, S.~{Arduini}, J.~. {Augu{\`e}res}, J.~{Brinchmann},
  R.~{Cole}, M.~{Cropper}, C.~{Dabin}, L.~{Duvet}, A.~{Ealet}, and et~al.
\newblock {Euclid Definition Study Report}.
\newblock {\em ArXiv e-prints}, October 2011.

\bibitem{Pozzetti:2016cch}
L.~Pozzetti, C.~M. Hirata, J.~E. Geach, A.~Cimatti, C.~Baugh, O.~Cucciati,
  A.~Merson, P.~Norberg, and D.~Shi.
\newblock {Modelling the number density of H$\alpha$ emitters for future
  spectroscopic near-IR space missions}.
\newblock {\em Astron. Astrophys.}, 590:A3, 2016.

\bibitem{Abdalla:2015zra}
Filipe~Batoni Abdalla et~al.
\newblock {Cosmology from HI galaxy surveys with the SKA}.
\newblock {\em PoS}, AASKA14:017, 2015.

\bibitem{Yahya:2014yva}
S.~Yahya, P.~Bull, M.~G. Santos, M.~Silva, R.~Maartens, P.~Okouma, and
  B.~Bassett.
\newblock {Cosmological performance of SKA HI galaxy surveys}.
\newblock {\em Mon. Not. Roy. Astron. Soc.}, 450:2251--2260, 2015.

\bibitem{Harrison:2016stv}
Ian Harrison, Stefano Camera, Joe Zuntz, and Michael~L. Brown.
\newblock {SKA Weak Lensing I: Cosmological Forecasts and the Power of
  Radio-Optical Cross-Correlations}.
\newblock 2016.

\bibitem{Bonaldi:2016lbd}
Anna Bonaldi, Ian Harrison, Stefano Camera, and Michael~L. Brown.
\newblock {SKA Weak Lensing II: Simulated Performance and SurveyDesign
  Considerations}.
\newblock 2016.

\bibitem{Camera:2016owj}
Stefano Camera, Ian Harrison, Anna Bonaldi, and Michael~L. Brown.
\newblock {SKA Weak Lensing III: Added Value of Multi-Wavelength Synergies for
  the Mitigation of Systematics}.
\newblock 2016.

\bibitem{Tyson:2003kb}
J.~Anthony Tyson.
\newblock {Large synoptic survey telescope: Overview}.
\newblock {\em Proc. SPIE Int. Soc. Opt. Eng.}, 4836:10--20, 2002.

\bibitem{Amanullah:2010vv}
R.~Amanullah et~al.
\newblock {Spectra and Light Curves of Six Type Ia Supernovae at 0.511 < z <
  1.12 and the Union2 Compilation}.
\newblock {\em Astrophys. J.}, 716:712--738, 2010.

\bibitem{2007PhRvL..98b1101K}
D.~J. {Kapner}, T.~S. {Cook}, E.~G. {Adelberger}, J.~H. {Gundlach}, B.~R.
  {Heckel}, C.~D. {Hoyle}, and H.~E. {Swanson}.
\newblock {Tests of the Gravitational Inverse-Square Law below the Dark-Energy
  Length Scale}.
\newblock {\em Physical Review Letters}, 98(2):021101, January 2007.

\bibitem{Dossett:2014oia}
Jason Dossett, Bin Hu, and David Parkinson.
\newblock {Constraining models of f(R) gravity with Planck and WiggleZ power
  spectrum data}.
\newblock {\em JCAP}, 1403:046, 2014.

\bibitem{Hojjati:2015ojt}
Alireza Hojjati, Aaron Plahn, Alex Zucca, Levon Pogosian, Philippe Brax,
  Anne-Christine Davis, and Gong-Bo Zhao.
\newblock {Searching for scalar gravitational interactions in current and
  future cosmological data}.
\newblock {\em Phys. Rev.}, D93(4):043531, 2016.

\end{thebibliography}

\end{document}